\documentclass[11pt,a4paper]{article}

\usepackage{amsmath,amsfonts,amssymb,amsthm,cite}

\evensidemargin=\oddsidemargin
\advance\topmargin by -\headheight
\advance\topmargin by -\headsep

\newtheorem{thm}{Theorem}[section]
\newtheorem{lem}[thm]{Lemma}
\newtheorem{cor}[thm]{Corollary}
\newtheorem{prop}[thm]{Proposition}
\theoremstyle{definition}
\newtheorem{defn}[thm]{Definition}
\newtheorem{cond}[thm]{Condition}
\theoremstyle{remark}
\newtheorem{rem}[thm]{Remark}

\numberwithin{equation}{section}        

\DeclareMathOperator{\dv}{div}          
\DeclareMathOperator{\Isom}{Isom}       
\DeclareMathOperator{\Ric}{Ric}         
\DeclareMathOperator{\spec}{sp}         
\DeclareMathOperator{\supp}{supp}       
\DeclareMathOperator{\Tr}{Tr}           

\renewcommand{\a}{\alpha}               
\newcommand{\B}{\mathcal{B}}            
\newcommand{\C}{\mathbb{C}}             
\newcommand{\Coo}{C^\infty}             
\newcommand{\del}{\partial}             
\newcommand{\Dslash}{{D\mkern-11.5mu/\,}} 
\newcommand{\eps}{\varepsilon}          
\newcommand{\Ga}{\Gamma}                
\renewcommand{\H}{\mathcal{H}}          
\newcommand{\half}{{\mathchoice{\thalf}{\thalf}{\shalf}{\shalf}}}
\newcommand{\hideqed}{\renewcommand{\qed}{}} 
\newcommand{\I}{\mathcal{I}}            
\renewcommand{\L}{\mathcal{L}}          
\newcommand{\la}{\lambda}               
\newcommand{\Mop}{\star}                
\newcommand{\N}{\mathbb{N}}             
\newcommand{\piso}[1]{\lfloor#1\rfloor} 
\newcommand{\R}{\mathbb{R}}             
\newcommand{\sepword}[1]{\quad\mbox{#1}\quad} 
\newcommand{\set}[1]{\{\,#1\,\}}        
\newcommand{\shalf}{{\scriptstyle\frac{1}{2}}} 
\newcommand{\slim}{\mathop{\mathrm{s\mbox{-}lim}}} 
\renewcommand{\SS}{\mathcal{S}}         
\newcommand{\T}{\mathbb{T}}             
\newcommand{\tG}{\widetilde{G}}         
\newcommand{\Th}{\Theta}                
\newcommand{\thalf}{\tfrac{1}{2}}       
\newcommand{\Trw}{\Tr_\omega}           
\newcommand{\x}{\times}                 
\newcommand{\Z}{\mathbb{Z}}             
\renewcommand{\.}{\cdot}                
\renewcommand{\:}{\colon}               

\newcommand{\comment}[1]{\relax}  

\begin{document}

\thispagestyle{empty}

\begin{center}

CENTRE DE PHYSIQUE TH\'EORIQUE$\,^1$\\
CNRS--Luminy, Case 907\\
13288 Marseille Cedex 9\\
FRANCE\\

\vspace{2cm}

{\Large
\textbf{Dixmier Traces on Noncompact Isospectral Deformations}} \\

\vspace{1cm}

{\large Victor Gayral,$^2$ Bruno Iochum,$^3$
and Joseph C. V\'arilly$^{4,5}$}\\

\vspace{1.5cm}

{\large\textbf{Abstract}}

\end{center}

\begin{quote}
We extend the isospectral deformations of Connes, Landi and
Dubois-Violette to the case of Riemannian spin manifolds carrying a
proper action of the noncompact abelian group $\R^l$. Under
deformation by a torus action, a standard formula relates Dixmier
traces of measurable operators to integrals of functions on the
manifold. We show that this relation persists for actions of $\R^l$,
under mild restrictions on the geometry of the manifold which
guarantee the Dixmier traceability of those operators.
\end{quote}

\vspace{1cm}


\vspace{1pc}
\noindent
CPT--2005/P.041\\
July 2005

\vspace{8pc}

\noindent $^1$ UMR 6207

-- Unit\'e Mixte de Recherche du CNRS et des
Universit\'es Aix-Marseille I, Aix-Marseille II et de l'Universit\'e
du Sud Toulon-Var

-- Laboratoire affili\'e \`a la FRUMAM -- FR 2291\\
$^2$ Matematisk Afdeling, Universitetsparken 5, 2100 K{\o}benhavn,
Denmark, gayral@math.ku.dk\\
$^3$ Also at Universit\'e de Provence, iochum@cpt.univ-mrs.fr\\
$^4$ Departamento de Matem\'aticas, Universidad de Costa Rica,
2060 San Jos\'e, Costa Rica\\
$^5$ Partially supported by European Commission grant
MKTD-CT-2004-509794.

\newpage


\section{Introduction}

The primary example of a noncommutative differential geometry is the
noncommutative torus \cite{ConnesTorus,RieffelRot}; its coordinate
algebra may be reconstructed from the algebra of smooth functions on
an ordinary torus $\T^l$ by deforming the product compatibly with the
rotation action of the torus, regarded as a compact abelian group, on
itself. The group $\T^l$ acts ergodically on the resulting deformed
algebra. Given a spin structure on $\T^l$, there is a Dirac operator
$\Dslash$ on the Hilbert space $\H$ of square-integrable spinors,
which is invariant under a lifted action of~$\T^l$; the deformed
algebra is also represented on this Hilbert space, giving rise to a
spectral triple \cite{Book} with the same Dirac operator: one speaks
of an \emph{isospectral deformation} of the triple
$(\Coo(\T^l),\H,\Dslash)$.

This example was generalized by Connes and Landi \cite{ConnesLa} to
the case of a $\T^l$-action, for $l \geq 2$, on a compact Riemannian
spin manifold. It was further refined by Connes and Dubois-Violette
\cite{ConnesDV} to encompass the case where the spin manifold need
not be compact but still carries a smooth torus action. In all such
cases, the Dirac operator interacts with the deformed algebra to
provide a isospectral deformation of the standard commutative
spectral triple.

Isospectral deformations arising from noncompact group actions provide
a more challenging analytic framework. It was established by Rieffel
\cite{RieffelDefQ} that Moyal deformations under actions of~$\R^l$
have good analytic properties, both at the level of $C^*$-algebras and
in terms of the smooth subalgebras for the action. This deformation
construction goes through when the symmetry group is abelian, so that
$\T^l$ and $\R^l$ are the cases of primary interest. However, the
compatibility of Moyal deformations with (invariant) Dirac operators
on noncompact spin manifolds poses additional issues for the
construction of deformed spectral triples. These issues have been
addressed and resolved in the ``flat'' case of the affine space
$\R^{2m}$ with translation action, whose deformations are Moyal
``planes'', in our \cite{Himalia} and in~\cite{GayralThese}.

In this paper, we consider proper actions of a connected abelian group
$\T^k \x \R^{l-k}$ on a (not necessarily compact) $n$-dimensional
Riemannian spin manifold $M$. This can be thought of as a proper,
hence free, action of $\R^{l-k}$ on a $\T^k$-twisted Connes--Landi
spectral triple; we therefore deal mainly with the subcase where $k =
0$. The detailed geometry of the manifold (isoperimetry, curvature
bounds) plays a role in establishing the heat-kernel properties of~$M$
and in determining the interplay of the isospectrally deformed algebra
with the Laplacian and the Dirac operator. From Connes' trace theorem
\cite{ConnesAction} for the case of compact manifolds, one expects
that operators such as $L_f |\Dslash|^{-n}$ or $L_f \Delta^{-n/2}$,
where $L_f$ denotes the deformed product by a function
$f \in \Coo_c(M)$, should lie in the Dixmier trace-class, and their
Dixmier traces should be proportional to the integral of $f$ with
respect to the Riemannian volume form. We show that this hope is
fulfilled in the noncompact case, under suitable conditions on the
geometry of~$M$. This general result was foreshadowed in the flat case
in~\cite{Himalia} and is extended here to a more general setting.

\vspace{6pt}

In Section~\ref{sec:Moyal-defn}, we review the Moyal products on
manifolds with an $\R^l$-action, to fix the notation. In
Section~\ref{sec:Hilb-anal}, we show that the Hilbert--Schmidt norm of
operators of the form $L_f\,h(\Dslash)$ is independent of the
deformation. In Section~\ref{sec:Schatten-est}, after discussing how
the required geometric properties yield bounds on the heat kernel, we
identify the Schatten classes $\L^p$ to which several such operators
belong, and show that certain important cases they lie in the weak
Schatten class $\L^{n,\infty}$, so that $L_f\,(1 + \Dslash^2)^{-n/2}$
belongs to the Dixmier trace class $\L^{1,\infty}$. The proof extends
and simplifies the argument of~\cite{Himalia}, based on Cwikel's
inequality~\cite{SimonTrace}.

In the final Sections \ref{sec:Trw-periodic}
and~\ref{sec:Trw-aperiodic}, we compute the desired Dixmier traces,
for both periodic and aperiodic actions of $\R^l$. In the aperiodic
case, the geometry is straightforward but the analysis is not, since
Dixmier traces, unlike ordinary integrals, do not admit monotone or
dominated convergence theorems: the heuristic extension of the compact
case put forward in \cite[Thm.~4.2]{Odysseus} is therefore unsupported
in general. We show, nevertheless, how to overcome this objection for
algebras arising from Moyal deformations.

\section{Moyal products on manifolds}
\label{sec:Moyal-defn}

\begin{defn}
Let $(M,g)$ be an $n$-dimensional (not necessarily compact) Riemannian
spin manifold which is geodesically complete and without boundary. Let
$G$ be a connected abelian Lie group of rank~$l$ (so that
$G \simeq \T^k \x \R^{l-k}$ for some $k = 0,1,\dots,l$), with
$l \geq 2$. Assume that $M$ is endowed with an isometric effective
action of~$G$, denoted $\a \: G \to \Isom(M,g)$, which is
\emph{smooth} (i.e., the map $G \x M \to M : (z,p) \mapsto \a_z(p)$ is
smooth) and \emph{proper}. Thus $M$ is a proper $G$-manifold in the
sense of~\cite{Michor}.
\end{defn}

For brevity, we often write $z\.p := \a_z(p)$. We also denote by $\a$
the induced action by automorphisms on $\Coo(M)$, i.e.,
$\a_z f(p) := f(\a_{-z}(p))$ for $p \in M$. Let $X_1,\dots,X_l$ be the
infinitesimal vector fields associated to the action, namely
$X_j(f) := \frac{\del}{\del z^j}(\a_z f)\bigr|_{z=0}$, for
$f \in \Coo(M)$.

Let $S \to M$ be the spinor bundle and $\H := L^2(M,S)$ be the
separable Hilbert space of its square integrable sections. Each
compactly supported smooth function $f \in \Coo_c(M)$ defines a
bounded operator $M_f$ on $\H$ by pointwise multiplication,
$M_f(\psi) := f\psi$.

The isometric action $\a$ lifts to $S$ modulo~$\pm 1$, as is pointed
out in \cite{ConnesDV}: for a suitable double covering
$p \: \tG \to G$, where $\tG$ is also isomorphic to
$\T^k \x \R^{l-k}$, we can find a group of unitary operators
$\set{V_{\tilde z} : \tilde z \in \tG}$ on $\H$ which covers the group
of isometries $\set{\a_z : z \in G}$ in the sense that
\begin{equation}
V_{\tilde z}(f\psi) = (\a_z f)\, V_{\tilde z}\psi,
\label{eq:spin-lift}
\end{equation}
whenever $\psi \in \H$, $f \in \Coo_c(M)$, and $p(\tilde z) = z$. In
general, unless $k = 0$, this spin lifting does not split: if
$p(\tilde z) = p(\tilde z')$ then $V_{\tilde z} = \pm V_{\tilde z'}$
but the sign cannot be taken globally to be $+1$. In what follows, we
shall ignore this nuance and shall suppose (in the notation) that the
spin lifting does split, writing $V_z$ instead of~$V_{\tilde z}$; thus
\eqref{eq:spin-lift} will be rewritten here as
$V_z(f\psi) = (\a_z f)\, V_z\psi$.

\begin{defn}
\label{df:Moyal-defn}
Let $\Th \in M_l(\R)$ be a fixed real skew-symmetric matrix.
For $f,h \in \Coo_c(M)$, the usual pointwise product of $f$ and $h$ can
be deformed by the group action $\a$, as follows~\cite{RieffelDefQ}:
\begin{equation}
f \Mop h := (2\pi)^{-l} \int_{\R^l}\int_{\R^l}
e^{-iy z} \, \a_{\half\Th y}(f) \, \a_{-z}(h) \,d^ly \,d^lz.
\label{eq:Moyal-defn}
\end{equation}
Thus, $\Mop$ is a bilinear product on $\Coo_c(M)$ with values in
$\Coo(M)$; its associativity can be checked directly.
\end{defn}

\begin{rem}
We could have written $\star_{_\Th}$ instead of $\Mop$, had we
needed to emphasize the dependence of the deformation on the
parameter matrix~$\Th$. When $\Th = 0$, the oscillatory
integral \eqref{eq:Moyal-defn} collapses to the usual pointwise
product of functions.

When $\Th$ is not invertible, the product \eqref{eq:Moyal-defn}
reduces to a twisted product associated to the action
$\sigma := \a\bigr|_{V^\bot}$, where $V$ is the nullspace of~$\Th$, as
in \cite[Prop.~2.7]{RieffelDefQ}. In what follows, we shall take $\Th$
to be a fixed \emph{invertible} matrix. In particular, this implies
that the rank $l$ {\it is even}.
\end{rem}

\begin{rem}
Herein $\Th$ is taken to be fixed, but this restriction is not forced:
it has been shown by one of us, together with Gracia-Bond\'{\i}a and
Ruiz Ruiz~\cite{Melpomene}, that Rieffel's approach is compatible with
some variable noncommutativity matrices $\Th(x)$. Giving such a $\Th$
determines a Poisson structure $\Pi_\Th$ on~$M$; and one expects to
find an associative star-product reproducing any given $\Pi_\Th$,
insofar as Kontsevich's formality theorem~\cite{Kontsevich} for
perturbative deformations carries over to the present context. Physics
would demand that this Poisson tensor should be a dynamical field,
interacting with the gravity background. Among the papers that already
invoke a variable~$\Th$, we may mention
\cite{AldrovandiSS,AschieriBDMSW,BehrS,CalmetK,DolanN,Selene,HashimotoT},
although most of the treatments so far have been kinematical.
\end{rem}

{}From now on, we shall treat separately $\T^l$-actions and effective
$\R^l$-actions, and we respectively talk about \emph{periodic} and
\emph{aperiodic} deformations; we obtain similar results in both
cases, albeit with different techniques. We do not deal directly with
mixed cases where the $\R^l$-action factors through an effective
action of $\T^k \x \R^{l-k}$ with $k = 1,\dots,l-1$; since the action 
of the toral and vectorial factors commute, one may reach the general 
case by composing a periodic and an aperiodic deformation.

In the aperiodic case, by the assumption of properness, the action is
also free: proper actions possess compact isotropy groups, but
$(\R^l,+)$ has no nontrivial compact subgroups. Under a proper, free
action, the orbit space $M/\R^l$ is a (Hausdorff) smooth manifold, and
the quotient map $\pi: M \to M/\R^l$ defines an $\R^l$-principal
bundle projection \cite[Thm.~1.11.4]{DuistermaatK}. Even though this
bundle is trivializable (see section \ref{sec:Trw-aperiodic}) and some
of our results could thereby be extracted from
\cite{Himalia,RennieSumm}, we adopt here an intrinsic approach more
compatible with eventual generalizations. In fact, the crucial 
Dixmier trace computation, in Theorem~\ref{th:Dix-tr} below, requires
new techniques.

In the periodic case, the action is obviously not free in general; in
\cite{Quijote}, one of us has shown that the set of singular points
for the action (i.e., points with nontrivial isotropy group) may give
rise to a new type of UV/IR mixing phenomenon for isospectral
deformations.

Note also that on noncompact manifolds, both periodic and aperiodic
deformations may occur; whereas when $M$ is compact, to be proper,
the action $\a$ must be periodic.

For torus actions, each $f \in \Coo_c(M)$ can be isotypically
decomposed via Peter--Weyl decomposition as a $\|.\|_\infty$-norm
convergent sequence (see \cite{ConnesLa, ConnesDV} for further
details):
\begin{equation}
f = \sum_{r\in\Z^l} f_r,
\label{eq:Peter-Weyl}
\end{equation}
where each homogeneous component $f_r$ satisfies
$\a_z(f_r) = e^{-iz r} f_r$, for all $z \in \T^l$. In this case, the
twisted product reproduces the canonical commutation relations for the
noncommutative $l$-torus, since
$$
f \Mop h = \sum_{r,s\in\Z^l} e^{-\half ir\.\Th s} \,f_r\,h_s.
$$
This computation shows that in the periodic noncompact case,
$(\Coo_c(M),\Mop)$ closes to an algebra: while this product is
nonlocal on the orbits of the action, the twisted product of two
functions $f,h \in \Coo_c(M)$ is again smooth and compactly supported
because $\supp f_r \subset \T^l\.(\supp f)$ and thus
$\supp(f \Mop h) \subset \T^l\.(\supp f \cap \supp h)$.

This need not be the case for aperiodic deformations, whose orbits are
noncompact. At this level of generality, one can only prove, using
Lemma~\ref{lm:sup} below, that
$f \Mop h \in \Coo(M) \cap L^\infty(M,\mu_g)$.

\section{Hilbertian analysis of deformed products}
\label{sec:Hilb-anal}

The Moyal product \eqref{eq:Moyal-defn} is defined on functions, but
the operator of left twisted multiplication $L_f\: h \mapsto f \Mop h$
may be lifted to spinors by replacing $\a_{-z} h$ by $V_{-z}\psi$ in
the defining formula. We find it convenient to work at both levels, on
the ``reduced'' Hilbert space of functions $\H_r := L^2(M,\mu_g)$ and
the ``full'' Hilbert space of square-integrable spinors
$\H = L^2(M,S)$. (Here $\mu_g$ is the Riemannian volume form.)
Somewhat abusively, we denote the left multiplication operators on
both spaces by the same symbol $L_f$, trusting that the context will
make clear which is which.

\subsection{Kernel properties of the $\Mop$ product}

We begin by showing that, for $f \in \Coo_c(M)$, the operator of left
twisted multiplication $L_f$ is a bounded kernel operator on $\H$ [or
on $\H_r$]. The same properties hold for the right twisted
multiplication operator $R_f$. We adopt the notation $M_f$ for the
(left or right) ordinary multiplication operator by $f$, corresponding
to the case $\Th = 0$.

\begin{defn}
\label{df:left-multn}
For $f \in \Coo_c(M)$, the operator of left twisted multiplication
$L_f$ acting on $\H = L^2(M,S)$ is defined for $p \in M$ by
\begin{equation}
L_f \psi(p) := (2\pi)^{-l} \int_{\R^l} \int_{\R^l} e^{-iy z}
(\a_{\half\Th y}f)(p) \,V_{-z}\psi(p) \,d^ly \,d^lz.
\label{eq:left-multn}
\end{equation}
(When the spin lifting of the action $\a$ does not split, the right
hand side must be replaced~by
$$
(2\pi)^{-l} \int_{\R^l} \int_{\widetilde{\R}^l} e^{-iy\,p(\tilde z)}
(\a_{\half\Th y}f)(p) \,V_{-\tilde z}\psi(p) \,d^ly \,d^l\tilde z,
$$
but we shall keep the version \eqref{eq:left-multn} to simplify the
notation.)
\end{defn}

\begin{defn}
\label{df:delta-g}
For any $p \in M$, let $\delta_p^g \in \mathcal{D}'(M)$ be the
distribution defined for $\phi \in \Coo_c(M)$ by
$$
\langle \delta_p^g, \phi \rangle
= \int_M \delta_p^g(p') \,\phi(p') \,\mu_g(p') := \phi(p).
$$
The distribution $\delta_p^g$ is represented in a local coordinate
system by $(\det g(x))^{-1/2} \,\delta(x - x')$, and the product
$\delta_p^g \,\mu_g$ can also be thought of as a de~Rham $n$-current
\cite{Schwartz}.
\end{defn}

\begin{prop}
\label{pr:kernel}
Let $\a$ be a smooth proper and isometric action of $\R^l$. When
$f \in \Coo_c(M)$, $L_f$ is a bounded kernel operator on $\H$ [or on
$\H_r$], with Schwartz kernel
\begin{equation}
K_{L_f}(p,p') = (2\pi)^{-l} \int_{\R^{2l}} e^{-iy.z}
f((-\half\Th y)\.p) \,\delta_{z\.p}^g(p') \,d^ly \,d^lz.
\label{eq:noyaux}
\end{equation}
\end{prop}

Before giving a proof, we need the following Lemma.

\begin{lem}
\label{lm:sup}
If $f \in \Coo_c(M)$ and the action $\a$ of $\R^l$ is free, then for
all $k \in \N$,
$$
\sup_{p\in M} \int_{\R^l} |\Delta_\a^k(\a_y f)(p)| \,d^ly < \infty,
$$
where $\Delta_\a := -\sum_{j=1}^l X_j^2$ is the Casimir operator.
\end{lem}

\begin{proof}
For any fixed $k$, the map
$\tilde f(p) := \int_{\R^l} |\Delta_\a^k(\a_y f)(p)| \,d^ly$ is well
defined since $\set{y \in \R^l : \a_y(p) \in \supp f}$ is compact for
each $p \in M$ \cite[p.~41]{Michor} because $f$ has compact support.
This gives rises to a finite $y$-integration and
$\tilde{f} \in \Coo(M)^G$. Let $\pi \: M \to M/\R^l$ be the projection
on the orbit space. Then $\tilde{f}$ factors through~$\pi$ to give a
map $\bar{f}$ defined by $\bar{f}(\pi(p)) := \tilde{f}(p)$. This
yields the result since $\bar{f} \in \Coo_c(M/{\R^l})$, because if
$p \notin \a_{\R^l}(\supp f)$, so that $\pi(p)$ is not in the compact
set $\pi(\supp f)$, then $\bar{f}(\pi(p)) = 0$.
\end{proof}

\begin{proof}[Proof of Proposition \ref{pr:kernel}]
For $\psi \in \H$, we can write, according to \eqref{eq:left-multn}
$$
L_f \psi(p) = (2\pi)^{-l} \int_{\R^l} \int_{\R^l} e^{-iy z}\,
\a_{\half\Th y}(f)(p) \int_M \delta^g_{z\.p}(p') \,\psi(p')
\,\mu_g(p') \,d^ly \,d^lz.
$$
The form of the kernel \eqref{eq:noyaux} is then obtained by
interchange of integrals. In the aperiodic case, that $\a$ is proper
is equivalent (see \cite[Defn.~5.1]{Michor}) to the compactness of
$\set{y \in \R^l : y\.K \cap L \neq \emptyset}$ for any compact
subsets $K$ and $L$ of $M$. So for $K = L = \{p\}$, for any $p \in M$,
this implies that its isotropy subgroup $H_p \subset \R^l$ is compact.
Hence $H_p = \{0\}$ for all $p \in M$ since $\a$ is free.

Boundedness of $L_f$ follows by a standard oscillatory-integral
trick \cite{Amalthea,Hormander,RieffelDefQ}:
\begin{align*}
L_f \psi(p)
&= (2\pi)^{-l} \int_{\R^l} \int_{\R^l} e^{-iy z} \,\a_{\half\Th y}(f)
\,V_{-z}\psi(p) \,d^ly \,d^lz
\\
&= (2\pi)^{-l} \int_{\R^l} (1 + |z|^2)^{-r} \int_{\R^l} (1 + |z|^2)^r
\,e^{-iy z} \,\a_{\half\Th y}(f) \,d^ly \,V_{-z}\psi(p) \,d^lz
\\
&= (2\pi)^{-l} \int_{\R^l} (1 + |z|^2)^{-r} \int_{\R^l}
((1 + \Delta_y)^r \,e^{-iy z}) \,\a_{\half\Th y}(f) \,d^ly
\,V_{-z}\psi(p) \,d^lz
\\
&= (2\pi)^{-l} \int_{\R^l} (1 + |z|^2)^{-r} \int_{\R^l} e^{-iy z}\,
((1 + \Delta_\a)^r \a_{\half\Th y}(f)) \,d^ly \,V_{-z}\psi(p) \,d^lz,
\end{align*}
where boundary terms vanish due to the compactness of $\supp f$.
Hence,
\begin{equation}
\|L_f \psi\| \leq (2\pi)^{-l} \|\psi\|\,
\biggl( \int_{\R^l} (1 + |z|^2)^{-r} \,d^lz \biggr) \sup_{p\in M}
\int_{\R^l} |(1 + \Delta_\a)^r \a_{\half\Th y}(f)(p)| \,d^ly
\label{eq:borne}
\end{equation}
is finite for $r > l/2$, thanks to Lemma~\ref{lm:sup}.
\end{proof}

\begin{rem}
In the periodic (compact or not) case, that $L_f$ is bounded for
$f \in \Coo_c(M)$ is a direct consequence of the Peter--Weyl
decomposition \eqref{eq:Peter-Weyl}. Indeed, the relation
\begin{equation}
L_{f_r} = M_{f_r} V_{-\half\Th r}
\label{eq:Lf-decomp}
\end{equation}
implies
$$
\|L_f\| \leq \sum_{r\in\Z^l} \|M_{f_r} V_{-\half\Th r}\|
\leq \sum_{r\in\Z^l} \|f_r\|_\infty,
$$
which is finite since the decomposition \eqref{eq:Peter-Weyl} is
convergent in the sup norm.

Furthermore, the Schwartz kernel of $L_f$ is
$$
K_{L_f}(p,p')
= \sum_{r\in\Z^l} f_r(p) \,\delta^g_{\half\Th r\.p}(p').
$$
\end{rem}

\begin{rem}
{}From the estimates used in the proof of Proposition~\ref{pr:kernel},
it is clear that this result, as well as all the statements of this
section, holds for a wider class of smooth functions decreasing fast
enough at infinity. It is in particular the case for smooth functions
which satisfy (with an obvious abuse of notation)
$\int_{\R^l} \a_y(f) \,d^ly \in \Coo_c(M/\R^l)$ and
$\int_{\R^l} |y^\beta\,X^\gamma\,\a_y(f)| \,d^ly < \infty$, for all
$\beta,\gamma \in \N^l$, i.e., which are compactly supported once
projected on the orbit space, and which are in the Schwartz space of
the orbits. However, for the sake of simplicity, we only consider in
the sequel functions in $\Coo_c(M)$, which of course have those
properties.
\end{rem}

\subsection{Hilbert--Schmidt norm invariance}
\label{sec:HS}

We are now concerned with invariance properties for kernels of
operators of type $h(\Dslash)$, where $h$ is any bounded positive
smooth function, and $\Dslash$ is the Dirac operator on~$S$. Since we
want $\Dslash$ to be essentially selfadjoint with domain $\Coo_c(M)$
(and we still denote by $\Dslash$ its selfadjoint closure), it is
sufficient, by a result of Wolf \cite{Wolf}, to assume from now on
that \emph{$M$ is geodesically complete}.

\begin{lem}
\label{lm:invkernel}
Let $h$ be a bounded positive smooth function on $\R$. Then the kernel
$K_{h(\Dslash)}$ is $\a$-invariant: for all $z \in \R^l$,
$p,p' \in M$,
$$
K_{h(\Dslash)}(z\.p, z\.p') = K_{h(\Dslash)}(p,p'),
$$
except possibly on a nullset of $M \x M$.
\end{lem}

\begin{proof}
This is a direct consequence of the isometry property of~$\a$; indeed,
the invariance of the Levi-Civita connection for~$g$ entails
invariance of the spin connection under the lifted action on spinors,
so that $V_z \Dslash V_{-z} = \Dslash$ for all~$z$.

This implies that $[V_z, h(\Dslash)] = 0$ for all $z \in \R^l$. Thus,
for $\psi \in \H$, the invariance of the Riemannian volume form under
the diffeomorphism $\a_{-z}$ yields
$$
\int_M K_{h(\Dslash)}(z\.p, z\.p') \,\psi(p') \,\mu_g(p')
= \int_M K_{h(\Dslash)}(z\.p, p') \,\psi((-z)\.p') \,\mu_g(p').
$$
The right hand side equals $(h(\Dslash) V_z\psi) (z\.p) =
(V_{-z} h(\Dslash) V_z\psi)(p) = (h(\Dslash)\psi)(p)$.

Thus, $K_{h(\Dslash)}(\a_z(\.),\a_z(\.))$ and $K_{h(\Dslash)}$
represent the same operator on~$\H$.
\end{proof}

The main result of this section is the following equality, which shows
that the Hilbert--Schmidt norm of $L_f\, h(\Dslash)$ is independent of
the deformation parameters in~$\Th$.

\begin{thm}
\label{th:HS-norm}
Let $f \in \Coo_c(M)$ and $h$ be a bounded positive function on~$\R$
such that $M_f\, h(\Dslash)$ is a Hilbert--Schmidt operator. Then the
operator $L_f\, h(\Dslash)$ is also Hilbert--Schmidt, with
$$
\|L_f\, h(\Dslash)\|_2 = \|M_f\, h(\Dslash)\|_2.
$$
\end{thm}

\begin{proof}
First, by Proposition \ref{pr:kernel}, one can compute the kernel of
$L_f\, h(\Dslash)$ in terms of $K_{h(\Dslash)}$:
\begin{align}
K_{L_f h(\Dslash)}(p,p')
&= \int_M K_{L_f }(p,q) K_{h(\Dslash)}(q,p') \,\mu_g(q)
\nonumber \\
&= (2\pi)^{-l} \int_M \int_{\R^{2l}} e^{-iy z}
f((-\half\Th y)\.p)\, \delta^g_{z\.p}(q)\, K_{h(\Dslash)}(q,p')
\,d^ly \,d^lz \,\mu_g(q)
\nonumber \\
&= (2\pi)^{-l} \int_{\R^{2l}} e^{-iy z} f((-\half\Th y)\.p)\,
K_{h(\Dslash)}(z\.p,p') \,d^ly \,d^lz.
\label{eq:kernel-comp}
\end{align}
Therefore,
\begin{align*}
& \|L_f\, h(\Dslash)\|_2^2
= \int_{M\x M} |K_{L_f h(\Dslash)}(p,p')|^2 \,\mu_g(p)\,\mu_g(p')
\\
&= (2\pi)^{-2l} \int_{M\x M} \int_{\R^{4l}} e^{i(y_1 z_1 - y_2 z_2)}
\bar{f}((-\half\Th y_1)\.p) \, f((-\half\Th y_2)\.p)
\\
&\qquad \x
\overline{K_{h(\Dslash)}}(z_1\.p, p') \, K_{h(\Dslash)}(z_2\.p, p')
\,d^ly_1 \,d^lz_1 \,d^ly_2 \,d^lz_2 \,\mu_g(p)\,\mu_g(p')
\\
&= (2\pi)^{-2l} \int_{M\x M} \int_{\R^{4l}} e^{i(y_1 z_1 - y_2 z_2)}
\bar{f}((-\half\Th y_1 - z_2)\.p) \, f((-\half\Th y_2 - z_2)\.p)
\\
&\qquad \x
\overline{K_{h(\Dslash)}}((z_1-z_2)\.p, (z_1-z_2)\.p') \,
K_{h(\Dslash)}(p, (z_1-z_2)\.p')
\,d^ly_1 \,d^lz_1 \,d^ly_2 \,d^lz_2 \,\mu_g(p)\,\mu_g(p'),
\end{align*}
where we used the invariance of~$\mu_g$ under the isometries
$p \mapsto (-z_2)\.p$ and $p' \mapsto (z_1 - z_2)\.p'$. Now by
Lemma~\ref{lm:invkernel}, using the translation
$z_1 \mapsto z_1 + z_2$, the last expression becomes
\begin{align*}
(2\pi)^{-2l} & \int_{M\x M} \int_{\R^{4l}}
e^{i(y_1 (z_1 + z_2) - y_2 z_2)} \, \bar{f}((-\half\Th y_1 - z_2)\.p)
\, f((-\half\Th y_2 - z_2)\.p)
\\
&\qquad\qquad \x
\overline{K_{h(\Dslash)}}(p, p') \, K_{h(\Dslash)}(p, z_1\.p')
\,d^ly_1 \,d^lz_1 \,d^ly_2 \,d^lz_2 \,\mu_g(p)\,\mu_g(p'),
\\
&= (2\pi)^{-2l} \int_{M\x M} \int_{\R^{4l}}
e^{i((y_1-2\Th^{-1}z_2) (z_1 + z_2) - y_2 z_2)} \,
\bar{f}((-\half\Th y_1)\.p) \, f((-\half\Th y_2)\.p)
\\
&\qquad\qquad \x
\overline{K_{h(\Dslash)}}(p, p') \, K_{h(\Dslash)}(p, z_1\.p')
\,d^ly_1 \,d^lz_1 \,d^ly_2 \,d^lz_2 \,\mu_g(p)\,\mu_g(p'),
\end{align*}
on making the translations $y_1 \mapsto y_1 - 2\Th^{-1}z_2$ and
$y_2 \mapsto y_2 - 2\Th^{-1}z_2$. This yields
\begin{align*}
(2\pi)^{-l} & \int_{M\x M} \int_{\R^{2l}} e^{iy z} \,
\bar{f}((-\half\Th y)\.p) \, f((-\half\Th y - z)\.p)
\\
&\qquad\qquad \x
\overline{K_{h(\Dslash)}}(p, p') \, K_{h(\Dslash)}(p, z\.p')
\,d^ly \,d^lz \,\mu_g(p)\,\mu_g(p')
\\
&= (2\pi)^{-l} \int_{M\x M} \int_{\R^{2l}} e^{iy z} \,
\bar{f}(p) \, f((-z)\.p) \,
\overline{K_{h(\Dslash)}}(p, p') \, K_{h(\Dslash)}(p, z\.p')
\,d^ly \,d^lz \,\mu_g(p)\,\mu_g(p')
\\
&= \int_{M\x M} |f(p)|^2 |K_{h(\Dslash)}(p,p')|^2
\,\mu_g(p)\,\mu_g(p') = \|M_f\, h(\Dslash)\|_2^2.
\end{align*}
The second equality uses the isometries
$p \mapsto (\half\Th y)\.p$ and $p' \mapsto (\half\Th y)\.p'$.
\end{proof}

\begin{rem}
Naturally, this result is still true in the restricted case of a
scalar Laplacian, i.e., for $L_f\,h(\Delta_r)$, with $\Delta_r$ the
scalar Laplacian acting on the reduced Hilbert space
$\H_r = L^2(M,\mu_g)$.

We shall see in the next section sufficient conditions on $h$
implying that $M_f\,h(\Dslash)$ lies in the Hilbert--Schmidt ideal.
\end{rem}

\begin{cor}
\label{cr:tr-invt}
If $L_f\,h(\Dslash)$ and $M_f\,h(\Dslash)$ are trace-class operators, 
then their traces coincide:
$$
\Tr(L_f\, h(\Dslash)) = \Tr(M_f\, h(\Dslash)).
$$
\end{cor}

\begin{proof}
The translation-invariance property of Lemma~\ref{lm:invkernel} and
the expression \eqref{eq:kernel-comp} for the kernel of
$L_f\, h(\Dslash)$ yield the equalities
\begin{align*}
\Tr(L_f h(\Dslash))
&= \int_M K_{L_f\,h(\Dslash)}(p,p) \,\mu_g(p)
\\
&= (2\pi)^{-l} \int_M \int_{\R^{2l}} e^{-iy z} f((-\half\Th y)\.p)\,
K_{h(\Dslash)}(z\.p,p) \,d^ly \,d^lz \,\mu_g(p)
\\
&= (2\pi)^{-l} \int_M \int_{\R^{2l}} e^{-iy z} f(p')\,
K_{h(\Dslash)}(z\.p',p') \,d^ly \,d^lz \,\mu_g(p')
\\
&= \int_M f(p') \,K_{h(\Dslash)}(p',p') \,\mu_g(p')
= \Tr(M_f h(\Dslash)).
\tag*{\qed}
\end{align*}
\hideqed
\end{proof}

The Riemannian volume form gives a natural trace for the twisted
product.

\begin{lem}
\label{lm:trace}
For $f,h \in \Coo_c(M)$,
$$
\int_M (f \Mop h) \,\mu_g = \int_M f \, h \, \mu_g.
$$
\end{lem}

\begin{proof}
It is enough to notice that, with $p \in M$,
\begin{align*}
\int_M f \Mop h(p) \,\mu_g(p)
&= (2\pi)^{-l} \int_M \int_{\R^{2l}} e^{-iy z} f((-\half\Th y)\.p)
\,h(z\.p) \,d^ly \,d^lz \,\mu_g(p)
\\
&= (2\pi)^{-l} \int_M \int_{\R^{2l}} e^{-iy z}
f((-\half\Th y - z)\.p) \,h(p) \,d^ly \,d^lz \,\mu_g(p)
\\
&= (2\pi)^{-l} \int_M \int_{\R^{2l}} e^{-iy z}
f((-z)\.p) \,h(p) \,d^ly \,d^lz \,\mu_g(p)
\\
&= \int_M f(p) \,h(p) \,\mu_g(p),
\end{align*}
using the isometry $p \mapsto (-z)\.p$ and the translation
$z \mapsto z - \half\Th y$.
\end{proof}

\begin{rem}
For formal deformations, Felder and Shoikhet~\cite{FelderSh} have shown
that a divergenceless Poisson bivector field yields a star-product
which is tracial. The divergence of $\Pi_\Th$ is a vector field, given in
local coordinates by
$$
\dv \Pi_\Th = (\del_j\Th^{ij} + \Ga_{lk}^l \Th^{ik}) \,\del_i;
$$
where $\Ga^i_{jk}$ are the Christoffel symbols for the metric~$g$.
Thus, $\Pi_\Th$ will be divergenceless if and only if
\cite[Chap.~7]{Polaris}:
$$
\Th^{ij} \,\del_i(\log\sqrt{\det g}) + \del_l \Th^{lj} = 0.
$$
This implies that $\Th$ must be of constant rank~\cite{FalcetoC}. A
result parallel to that of~\cite{FelderSh}, in our context, would
suggest that variable noncommutativity matrices should prevail in
non-flat backgrounds, although one may admit a nonconstant,
divergenceless $\Th$ in a flat background (the case considered
in~\cite{Melpomene}) or a constant, divergenceless $\Th$ in a non-flat
background (as we do here).
\end{rem}

In what follows, we shall take advantage of the possibility of viewing
$L_f$, for $f \in \Coo_c(M)$, as an integral of bounded operators:
\begin{equation}
\label{eq:int-pres}
L_f = (2\pi)^{-l} \int_{\R^{2l}} e^{-iy z} \,
V_{\half\Th y} \,M_f\, V_{-\half\Th y-z} \,d^ly \,d^lz.
\end{equation}
This is not a Bochner integral (the integral of the norm of the
integrand is not absolutely convergent), but rather a $\B(\H)$-valued
oscillatory integral, as shown in the proof of
Proposition~\ref{pr:kernel}.

The invariance property of the Hilbert--Schmidt norm can be
generalized as follows. One can construct left and right twists for a
wider class of bounded operators. For $A\in\B(\H)$ we formally
define its left and right twists by
\begin{align*}
L_A &:= (2\pi)^{-l} \int_{\R^{2l}} e^{-iy z} \,
V_{\half\Th y} \,A\, V_{-\half\Th y-z} \,d^ly \,d^lz \, ,
\\
R_A &:= (2\pi)^{-l} \int_{\R^{2l}} e^{-iy z} \,
V_{-z} \,A\, V_{z+\half\Th y} \,d^ly \,d^lz \, .
\end{align*}
These expressions are well defined, at least, for Hilbert--Schmidt
operators thanks to the following generalization of Theorem
\ref{th:HS-norm}.

\begin{thm}
Let $A$ be a Hilbert--Schmidt operator. Then $L_A$ and $R_A$
are also Hilbert--Schmidt operators and
$$
\|L_A\|_2 = \|R_A\|_2 = \|A\|_2 \, .
$$
\end{thm}

\begin{proof}
We treat $L_A$ only. The kernel $K_A$ of $A$ lies in
$L^2(M \x M, \mu_g \x \mu_g)$, and we can express $K_{L_A}$ in terms
of~$K_A$:
$$
K_{L_A}(p,p') = (2\pi)^{-l}\int_{\R^{2l}}
e^{-iy.z}\, K_A(\half\Th y\.p, z\.p') \, d^ly\,d^lz \, .
$$
Thus, routine computations yield that the map $K_A \mapsto K_{L_A}$ is
an isometry on $L^2(M \x M, \mu_g \x \mu_g)$:
$$
\|L_A\|_2 = \int_{M\x M} |K_{L_A}(p,p')|^2 \,\mu_g(p) \,\mu_g(p')
= \|A\|_2 \, .
\eqno \qed
$$
\hideqed
\end{proof}

\section{Schatten-class estimates for twisted multiplication operators}
\label{sec:Schatten-est}

In this section, we give Schatten-norm estimates for the operators
$M_f(1 + \Delta_r)^{-k}$ and $L_f(1 + \Delta_r)^{-k}$ acting on the
reduced Hilbert space $\H_r = L^2(M,\mu_g)$, where $\Delta_r$ is the
Laplacian $(d + d^*)^2$ reduced to 0-forms (in our convention, the
Laplacian is a positive operator). This will be done using heat kernel
estimates and the Laplace transform for $(1 + \Delta_r)^{-k}$, together
with Proposition~\ref{pr:kernel} and Theorem~\ref{th:HS-norm}. For
convenience and when no ambiguity can occur, we shall omit the
subscript $r$ for the reduced Laplacian.

We use the notations $\L^p(\H)$, $p \geq 1$, for the $p$-Schatten
class of operators on the Hilbert space $\H$ and $\L^{n,\infty}(\H)$
for the $n^+$-summable operators on $\H$.

\subsection{Some heat-kernel estimates}

We now need to make some more precise assumptions on the geometry
of~$M$, which give some (mild) controls on the asymptotics of the heat
kernel.

Let $K_t(p,p')$ denote the heat kernel, associated to the operator
$e^{-t\Delta_r}$, defined on $M \x M$ for $0 < t < \infty$. Recall
that, in full generality, each $K_t(p,p')$ is a smooth strictly
positive symmetric function on $M \x M$
\cite[Thm.~5.2.1]{DaviesHeatKer}.

\vspace{6pt}

For the remainder of the article, we shall suppose that the manifold
$M$ satisfies the following hypothesis.

\begin{cond} 
\label{cn:heat-ker}
$M$ is a complete connected Riemannian spin manifold of dimension
$n \geq 2$ without boundary such that
\begin{align}
\sup_{p\in M} \,\int_0^\infty t^k \,e^{-t} \,K_t(p,p) \,dt
&< \infty  \sepword{for all}  k > \frac{n}{2} - 1,
\label{eq:kernel1}
\\
\intertext{and for some $c > 0$,}
\sup_{p\in M} \,\int_m^\infty t^{-\half}\, e^{-t} \,K_t(p,p) \,dt
&< c\, m^{-(n-1)/2}  \sepword{for all}  m \in (0,1].
\label{eq:kernel2}
\end{align}
\end{cond}

These constraints imply a control of the heat kernel near $0$ and
$\infty$ which is sufficient for the Dixmier trace computations. They
are not too severe, as the next Lemma shows. Some such controls are
necessary because for any complete Riemannian manifold of finite
volume $V(M)$, in particular for a compact manifold,
$\int_1^\infty K_t(p,p) \,dt = \infty$ holds since
$\lim_{t\to\infty} K_t(p,p') = V(M)^{-1}$.

Let $B(p,r) := \set{p'\in M : d_g(p,p') < r}$ denote the geodesic
ball centered at~$p$ with radius $r$. The isoperimetric constant
$\I(M)$ is given \cite[p.~96]{Chavel} by
$$
\I(M) := \inf_\Omega \frac{A(\del\Omega)^n}{V(\Omega)^{n-1}},
$$
where $\Omega$ ranges over all open submanifolds with compact closure
in $M$ and with smooth boundary, $V(\Omega)$ and $A(\del\Omega)$ are
the Riemannian volume and area of $\Omega$ and $\del\Omega$
respectively.

\begin{lem}
\label{lm:good-geom}
Let $M$ be a complete Riemannian manifold satisfying one of the
following:
\begin{enumerate}
\item 
$M$ has Ricci curvature bounded from below, that is,
$\Ric(p) \geq (n - 1)\,\beta$, for all $p \in M$ and some
constant~$\beta$. Moreover,
$\sup_{p\in M} V(B(p,a))^{-1} < \infty$ for some $a > 0$.
\item
$M$ is noncompact with a positive injectivity radius, and there
exists $a > 0$ such that
$\sup_{p\in M} \I(B(p,a))^{-1} < \infty$. (This last property holds
if $M$ has a positive isoperimetric constant: $\I(M) > 0$.)
\end{enumerate}
Then the inequalities \eqref{eq:kernel1} and \eqref{eq:kernel2} hold
for~$M$.
\end{lem}

\begin{proof}
Assume the first condition. In \cite[Lemma 15]{DaviesBound} ---see
also \cite{DaviesHeatKer}--- we get the following estimates. Given
$\eps > 0$, there exists a constant $c_\eps$ such that, for all
$t > 0$ and $p \in M$,
\begin{equation*}
0 \leq K_t(p,p) \leq c_\eps(n)\,V(B(p,t^{1/2}))^{-1} \,e^{(\eps-E)t},
\end{equation*}
where $E := \inf\spec(\Delta) \geq 0$. Since, by
\cite[Prop.~4.1]{CheegerGT},
$$
V(B(p,r)) \geq c\,r^n \,V(B(p,1))  \sepword{for} 0 < r < 1,
$$
we get
\begin{equation}
\label{eq:k2}
K_t(p,p) \leq
\begin{cases}
C_2(\eps)\, t^{-n/2}\, V(B(p,1))^{-1}\, e^{(\eps-E)t}, & t \leq 1,
\\[3pt]
C_3(\eps)\, V(B(p,1))^{-1}\, e^{(\eps-E)t},            & t > 1.
\end{cases}
\end{equation}
Now suppose instead that the second condition holds. In \cite[Thm.~8,
p.~198]{Chavel}, it is proved that the heat kernel has an upper bound:
for all $p \in M$ and $r > 0$ for which $\overline{B(p,r)}$ lies in
the image of the exponential map $\exp_p$, the following estimate
holds:
\begin{equation}
\label{eq:k1}
K_t(p,p) \leq C_1(n)\, (t^{-n/2} + r^{-(n+2)}\,t) \, \I(B(p,r))^{-1}.
\end{equation}
In Case 1, we assumed that $\sup_{p\in M} V(B(p,a))^{-1} < \infty$ for
some $a > 0$. Similarly, the constraint on the injectivity radius in
Case~2 implies that for some $r_0$, $\overline{B(p,r_0)}$ lies in the
image of the exponential maps $\exp_p$ for all $p \in M$.

Thus the estimates \eqref{eq:k1} and \eqref{eq:k2} yield
\begin{equation*}
K_t(p,p)
\leq c_1\,(t^{-n/2} + c_2\,t) \max(e^{(\eps_0-E)t}, 1)
\end{equation*}
for some positive constants $c_1$, $c_2$, independent of $p$,
 for a fixed $\eps = \eps_0 < 1$.

Let $b = \max(\eps_0-E-1,-1)$. Then $b<0$ and
\begin{align*}
\sup_{p\in M} \,\int_0^{\infty} t^k \,e^{-t} \,K_t(p,p) \,dt
&\leq c_1\, \int_0^{\infty} t^{k-\frac{n}{2}} \, e^{bt} \,dt
+ c_2 \, \int_0^{\infty} t^{k+1}\, e^{bt} \,dt
\\
&= c_1 \, \Ga(k - \tfrac{n}{2} + 1)\, b^{-(k-\frac{n}{2}+1)}
+ c_2\, \Ga(k + 2) \, b^{-(k+2)}
\end{align*}
is finite and \eqref{eq:kernel1} holds.

Similarly, we get
\begin{align*}
\sup_{p\in M} \, \int_m^\infty t^{-\half} \,e^{-t} \,K_t(p,p) \,dt
& \leq c_1 \, \sup_{p\in M} \, \int_m^\infty t^{-(n+1)/2} \,dt
+ c_2 \sup_{p\in M} \, \int_m^\infty t \,e^{bt} \,dt
\\
&= c_1 \, \frac{2}{n - 1}\, m^{-(n-1)/2}
+ c_2 \, \frac{1 - mb}{b^2} \, e^{mb}.
\end{align*}
Since $(1 - mb) e^{mb} < 1 - mb < (1 - mb)\,m^{-(n-1)/2}$ for
$0 < m < 1$, the inequality \eqref{eq:kernel2} also holds.
\end{proof}

\begin{rem}
Since $\sup_{p\in M} K_t(p,p)$ is decreasing in $t$, the condition
\eqref{eq:kernel1} is satisfied if, for some $c' > 0$,
$$
\sup_{p\in M} K_t(p,p) < c'\,e^t\,t^{-n/2} \sepword{for all} 0 < t < 1.
$$
It is known (see \cite{CoulhonIsop}, for instance) that
$$
\|e^{-t\Delta}\|_{1\to\infty} = \sup_{p\in M} K_t(p,p).
$$
Thus, changing $\Delta$ to $1 + \Delta$, the condition
\eqref{eq:kernel1} is guaranteed by
\begin{equation}
\|e^{-t(1+\Delta)}\|_{1\to\infty} < c' \,t^{-n/2}
\sepword{for all} 0 < t < 1.
\label{eq:contrainte}
\end{equation}
This can be reformulated in many different ways, according 
to~\cite{CoulhonDim}. For $n > 2$, \eqref{eq:contrainte} is 
equivalent to the boundedness of the operator
$(1 + \Delta)^{-1/2} : L^2(M,\mu) \to L^{2n/(n-2)}(M,\mu)$, or of the 
operator $(1 + \Delta)^{-\a/2} : L^p(M,\mu) \to L^q(M,\mu)$, for
$1 < p < \infty$, $\alpha p < n$ and
$\tfrac{1}{q} = \tfrac{1}{p} - \tfrac{\a}{n}$. This can be used in the
next subsection.

According to \cite[Prop.~1.2]{CoulhonIsop}, this implies that
$\sup_{p\in M} V(B(p,1))^{-1} < \infty$, for $n > 2$.

Note that a strictly positive isoperimetric constant is a stronger
condition than \eqref{eq:kernel1}: see~\cite{CoulhonIsop}. For
instance, when $M = \R^n$ with its Euclidean metric,
$K_t(p,p) = (4\pi\,t)^{-n/2}$ for all $p \in M$ and $t > 0$.
\end{rem}

\begin{rem}
\label{rk:bdd-geom}
Recall that a \emph{bounded geometry} on a connected manifold $M$ is a
Riemannian metric on~$M$ whose injectivity radius is positive and
satisfies $|\nabla^k R| \leq C_k$, $k \in \N$, i.e., every covariant
derivative of the Riemann curvature tensor is bounded: see
\cite{Chavel,Shubin,Triebel}. Such a Riemannian manifold is
automatically complete and satisfies the Condition~\ref{cn:heat-ker}.
In fact, any $n$-dimensional manifold with positive injectivity radius
and Ricci curvature uniformly bounded below obeys an upper bound:
$\sup_{p\in M} K_t(p,p) \leq C \max(t^{-n/2},t^{-1/2})$ for all
$t > 0$: see \cite[Thm.~7.9]{Grigoryan}. Thus \eqref{eq:kernel1} and
\eqref{eq:kernel2} are valid.

Examples of manifolds with bounded geometry are given by Lie groups,
homogeneous manifolds with invariant metrics, covering manifolds of
compact manifolds with the lifted Riemannian metric, leaves of a
foliation on a compact manifold with a metric induced by the
Riemannian metric on the compact manifold. In particular, all
manifolds with a transitive group of isometries have
$C^\infty$-bounded geometry.
\end{rem}

\subsection{Schatten-class estimates}

We start with a straightforward consequence of~\eqref{eq:kernel1}.

\begin{lem}
\label{lm:gro}
Assume that $M$ satisfies Condition~\ref{cn:heat-ker}. Then
$(1 + \Delta)^{-k}$ is a bounded operator from $L^2(M,\mu_g)$ to
$L^\infty(M,\mu_g)$, for all $k > n/4$.
\end{lem}

\begin{proof}
Let $\phi \in L^2(M,\mu_g)$. Using the Cauchy--Schwarz inequality, 
positivity and symmetry of $K_{(1 + \Delta)^{-k}}$, positivity
of $\mu_g$, the product rule for kernel operators and the Laplace
transform $(1 + \Delta)^{-2k} =
\Ga(2k)^{-1} \int_0^\infty t^{2k-1}\,e^{-t\,(1 + \Delta)} \,dt$,
we get
\begin{align*}
\|(1 + \Delta)^{-k}\phi\|_\infty^2
&=
\sup_{p\in M} \biggl| \int_M K_{(1 + \Delta)^{-k}}(p,p') \,\phi(p')
\,\mu_g(p') \biggr|^2
\\
&\leq \|\phi\|_2^2 \, \sup_{p\in M} \int_M
|K_{(1 + \Delta)^{-k}}(p,p')|^2 \,\mu_g(p')
\\
&= \|\phi\|_2^2 \, \sup_{p\in M} \int_M
K_{(1 + \Delta)^{-k}}(p,p')\, K_{(1 + \Delta)^{-k}}(p',p) \,\mu_g(p')
\\
&= \|\phi\|_2^2 \, \sup_{p\in M} K_{(1 + \Delta)^{-2k}}(p,p)
\\
&= \frac{\|\phi\|_2^2}{\Ga(2k)} \,
\sup_{p\in M} \int_0^\infty t^{2k-1} \,e^{-t} \,K_t(p,p) \,dt.
\end{align*}
By \eqref{eq:kernel1}, the $t$-integral is finite when
$k > \frac{n}{4}$, so
$\|(1 + \Delta)^{-k} \phi\|_\infty \leq c(k)\, \|\phi\|_2$.
\end{proof}

We now give the principal result of this subsection.
Condition~\ref{cn:heat-ker} is assumed throughout.

\begin{prop}
\label{pr:HiSc}
For $f\in L^2(M,\mu_g)$, the operator $M_f\,(1 + \Delta)^{-k}$ is
Hilbert--Schmidt for $k > n/4$ and satisfies
$$
\|M_f\,(1 + \Delta)^{-k}\|_2 \leq C_k(n)\,\|f\|_2.
$$
\end{prop}

\begin{proof}
That the operator $M_f\,(1 + \Delta)^{-k}$ is Hilbert--Schmidt is a
consequence of the factorization principle of Grothendieck ---see
\cite[Ex.~11.18]{Defant}, for instance--- which is this context says 
that when two operators $B\: L^2(X,\mu) \to L^\infty(X,\mu)$ and
$A\: L^\infty(X,\mu) \to L^2(X,\mu)$ are both bounded, their product
$AB$ is a Hilbert--Schmidt operator on $L^2(X,\mu)$.

Since for $f \in L^2(M,\mu_g)$, $M_f$ is bounded from
$L^\infty(M,\mu_g)$ into $L^2(M,\mu_g)$, Lemma~\ref{lm:gro} shows that
$M_f\,(1 + \Delta)^{-k}$ is Hilbert--Schmidt for $k > n/4$. For the
Hilbert--Schmidt-norm estimate, we again use \eqref{eq:k1},
\eqref{eq:k2} and Laplace transform techniques:
\begin{align*}
\|M_f\,(1 + \Delta)^{-k}\|_2^2
&= \int_{M\x M} |f(p)|^2 |K_{(1 + \Delta)^{-k}}(p,p')|^2
\,\mu_g(p) \,\mu_g(p')
\\
&= \int_M |f(p)|^2 \, K_{(1 + \Delta)^{-2k}}(p,p) \,\mu_g(p)
\\
&= \frac{1}{\Ga(2k)} \int_M |f(p)|^2 \,\mu_g(p)
\int_0^\infty t^{2k-1} \,e^{-t} \, K_t(p,p) \,dt
\\
&\leq C_k(n)^2 \,\|f\|_2^2,
\end{align*}
where we used  again \eqref{eq:kernel1}, the symmetry of
$K_{(1 + \Delta)^{-k}}$ and the product rule for kernels.
\end{proof}

\begin{rem}
The result of the previous Proposition can be generalized at least for
operators $M_f\,h(\sqrt{\Delta})$ where $h$ is a Laplace transform of
some function which behaves as $t^{k-1}$ when $t \downarrow 0$, for
$k > n/4$, and has fast enough decrease at infinity.
\end{rem}

\begin{thm}
\label{th:inter}
If $f \in L^p(M,\mu_g)$ with $2 \leq p < \infty$,
then $M_f\,(1 + \Delta)^{-k} \in \L^p(\H_r)$ for $k > n/4$.
\end{thm}

\begin{proof}
The case $p = 2$ is Proposition~\ref{pr:HiSc}. For $p = \infty$, we
use
$$
\|M_f\,(1 + \Delta)^{-k}\| \leq \|M_f\| \,\|(1 + \Delta)^{-k}\|
\leq \|f\|_\infty.
$$
We use complex interpolation for $2 < p < \infty$. Firstly, note that
we can always assume $f$ to be nonnegative, since
$$
\|M_f\| = \|M_{|f|}\|,  \qquad
\|M_f\,(1 + \Delta)^{-k}\|_2 = \|M_{|f|}\,(1 + \Delta)^{-k}\|_2.
$$
Then, for $f \geq 0$ in $L^p(M,\mu_g)$, we define the map
$$
F_p : z \mapsto M_f^{pz} \,(1 + \Delta)^{-kpz},
$$
for all $z$ in the strip
$S := \set{z \in \C : 0 \leq \Re z \leq \half}$. For all $y \in \R$,
$F_p(iy) = M_f^{ipy}\,(1 + \Delta)^{ikpy}$ is bounded with
$\|F_p(iy)\| \leq 1$; and for $z = \half + iy$,
Proposition~\ref{pr:HiSc} shows that
$$
\|F_p(\half + iy)\|_2 = \|M_f^{p/2}\,(1 + \Delta)^{-kp/2}\|_2
\leq C_{kp/2}(n)\, \|f^{p/2}\|_2 = C_{kp/2}(n)\, \|f\|_p^{p/2},
$$
which is finite because $k > n/2p$. Then, interpolation
\cite{SimonTrace} yields $F_p(z) \in \L^{1/\Re z}(\H_r)$ for all
$z \in S$, and
\begin{align*}
\|F_p(z)\|_{1/\Re z}
&\leq \|F_p(0)\|_\infty^{1-2\Re z}\, \|F_p(\half)\|_2^{2\Re z}
\leq \|M_f^{p/2}(1 + \Delta)^{-kp/2}\|_2^{2\Re z}
\\
&\leq C_{kp/2}(n)^{2\Re z}\, \|f^{p/2}\|_2^{2\Re z}
= C_{kp/2}(n)^{2\Re z}\, \|f\|_p^{p\Re z}.
\end{align*}
So, for $z = 1/p$, we get
$$
\|F_p(1/p)\|_p = \|M_f\,(1 + \Delta)^{-k}\|_p
\leq C_{kp/2}(n)^{2/p}\, \|f\|_p,
$$
and the result follows.
\end{proof}

\begin{prop}
\label{pr:interpolation}
Let $2 \leq p < \infty$ and $f \in \Coo_c(M)$. Then, if $\a$ is an
isometric proper action of $\R^l$ on $M$,
$L_f\,(1 + \Delta)^{-k} \in \L^p(\H_r)$ for all $k > n/2p$.
\end{prop}

\begin{proof}
The proof is essentially the same as the previous one, so we only
sketch it. Theorem~\ref{th:HS-norm} and Proposition~\ref{pr:HiSc}
imply that, for $k > n/4$,
$$
\|L_f\,(1 + \Delta)^{-k}\|_2 = \|M_f\,(1 + \Delta)^{-k}\|_2
\leq C_k(n)\, \|f\|_2.
$$
Moreover, by \eqref{eq:borne},
$$
\|L_f\,(1 + \Delta)^{-k}\| \leq \|L_f\| \leq \widetilde C_r(l) \,
\sup_{p\in M} \int_{\R^l} |(1 + \Delta_y)^r \a_{\half\Th y}f(p)| \,d^ly
=: \omega(f;r,l,n),
$$
which is finite whenever $r > l/2$. Defining
$G_p(z) := L_f\,(1 + \Delta)^{-kpz}$ for $z \in S$ and $k > n/2p$, we
conclude that, for all $y \in \R$,
$$
\|G_p(iy)\| = \|L_f\,(1 + \Delta)^{-ikpy}\| \leq \omega(f;r,l,n),
$$
and
$$
\|G_p(\half + iy)\|_2 = \|L_f\,(1 + \Delta)^{-kp/2}\|_2
\leq C_{kp/2}(n)\, \|f\|_2.
$$
Again, complex interpolation gives the result:
\begin{align}
\|L_f\,(1 + \Delta)^{-k}\|_p
&= \|G_p(p^{-1})\|_p
\leq \|G_p(0)\|_\infty^{1-2/p}\, \|G_p(2^{-1})\|_2^{2/p}
\notag \\
&\leq \omega(f;r,l,n)^{1-2/p}\, C_{kp/2}(n)^{2/p}\, \|f\|_2^{2/p}.
\tag*{\qed}
\end{align}
\hideqed
\end{proof}

\begin{rem}
Using again the Peter--Weyl decomposition \eqref{eq:Peter-Weyl}, one can
show that this interpolation result holds also for periodic noncompact
deformations.
\end{rem}

We now show that the previous Proposition extends directly to the
spinor bundle.

\begin{cond}
\label{cn:bdd-geom}
Assume from now on that $M$ satisfies Condition~\ref{cn:heat-ker}
and moreover has bounded scalar curvature.

This condition is satisfied for bounded geometries as noticed in
Remark~\ref{rk:bdd-geom}.
\end{cond}

\begin{cor}
\label{cl:interpolation}
Let $2 \leq p < \infty$ and $f \in \Coo_c(M)$. If $\a$ is an isometric
proper action of $\R^l$ on $M$, then $L_f\,(1 + \Dslash^2)^{-k}$ and
$L_f\,(1 + |\Dslash|)^{-2k}$ are in $\L^p(\H)$ for all $k > n/2p$.
\end{cor}

\begin{proof}
Since $(1 + \Dslash^2)^{k}(1 + |\Dslash|)^{-2k}$ is bounded, it
suffices to consider $L_f\,(1 + \Dslash^2)^{-k}$ only. For this
operator, the result follows from a simple comparison argument using
the Lichnerowicz formula
\begin{equation}
\Dslash^2 = \Delta + \tfrac{1}{4} R,
\label{eq:Lich-formula}
\end{equation}
where $R$ is the scalar curvature, bounded by hypothesis. Thus, the
result follows from
$$
(1 + \Dslash^2)^{-1}
= (1 + \Delta)^{-1} (1 - \tfrac{1}{4} R(1 + \Dslash^2)^{-1}).
\eqno \qed
$$
\hideqed
\end{proof}

Before finishing this subsection, we show for later use that the
following commutators have the same summability properties as
$L_f(1 + \Dslash^2)^{-k}$.

\begin{lem}
\label{lem:plenty}
If $f \in \Coo_c(M)$ and $2 \leq p < \infty$, then the operators
\begin{align*}
& [\Dslash, L_f]\,(1 + \Dslash^2)^{-k},
&& [|\Dslash|, L_f]\,(1 + \Dslash^2)^{-k},
&& [(1 + \Dslash^2)^{\half}, L_f]\,(1 + \Dslash^2)^{-k},
\\
& [\Dslash, L_f]\,(1 + |\Dslash|)^{-2k},
&& [|\Dslash|, L_f]\,(1 + |\Dslash|)^{-2k},
&& [(1 + \Dslash^2)^{\half}, L_f]\,(1 + |\Dslash|)^{-2k}
\end{align*}
all lie in $\L^p(\H)$ whenever $k > n/2p$.
\end{lem}

\begin{proof}
It is enough to prove this Lemma in the $(1 + \Dslash^2)^{-k}$ case.

For $[\Dslash,L_f](1 + \Dslash^2)^{-k}$, this is a direct consequence of
the isometry property of the action: since $\Dslash$ commutes with
(the lift to the spinor bundle of) the action, we obtain
$$
[\Dslash, L_f] = L_{[\Dslash, M_f]} = L_{\Dslash f},
$$
Hence the proof of Proposition \ref{pr:interpolation} applies with
$\Dslash f$ instead of $f$ because $\Dslash f \in \Coo_c(M)$.

For $[|\Dslash|,L_f]$, we can reduce the proof to the previous case
by using the following spectral identity for a positive operator $A$:
\begin{equation}
A = \frac{1}{\pi} \int_0^\infty \frac{A^2}{A^2 + \la}
\frac{d\la}{\sqrt{\la}}\ .
\label{eq:spectral-sqrt}
\end{equation}
Thus, for any positive number $\rho$,
\begin{align*}
& [|\Dslash|,L_f] = [|\Dslash|+\rho, L_f]
= \frac{1}{\pi} \int_0^\infty \frac{1}{(|\Dslash|+\rho)^2 + \la}
\bigl[ (|\Dslash|+\rho)^2, L_f \bigr]
\frac{1}{(|\Dslash|+\rho)^2 + \la} \,\sqrt{\la}\,d\la
\\
&= \frac{1}{\pi} \int_0^\infty \frac{1}{(|\Dslash|+\rho)^2 + \la}
\big(\Dslash[\Dslash, L_f] + [\Dslash, L_f]\Dslash
 + 2\rho |\Dslash| L_f - 2\rho L_f|\Dslash|\big)
\frac{1}{(|\Dslash|+\rho)^2 + \la} \,\sqrt{\la}\,d\la.
\end{align*}
Let us consider the different terms: since
$[\Dslash,L_f] = L_{\Dslash f}$, they are all of the same order in
$\Dslash$; we treat in detail only the first term since the proof goes
along the same lines for the others.

Commuting $[\Dslash,L_f]$ with the factor
$((|\Dslash|+\rho)^2 + \la)^{-1}$ to its right, the first term of the
last display equals:
\begin{align*}
&\frac{1}{\pi} \int_0^\infty
\frac{|\Dslash|+\rho}{((|\Dslash|+\rho)^2 + \la)^2}
\,\sqrt{\la}\,d\la\
\frac{\Dslash}{|\Dslash|+\rho} \,[\Dslash,L_f]
\\
&\qquad + \frac{1}{\pi} \int_0^\infty
\frac{1}{((|\Dslash|+\rho)^2 + \la)^2}
\Dslash \bigl[ (|\Dslash| + \rho)^2, [\Dslash,L_f] \bigr]
\frac{1}{(|\Dslash|+\rho)^2 + \la} \,\sqrt{\la}\,d\la
\\
&= \frac{1}{2} \frac{\Dslash}{|\Dslash|+\rho} \,[\Dslash,L_f]
 + \frac{1}{\pi} \int_0^\infty \frac{1}{((|\Dslash|+\rho)^2 + \la)^2}
\Dslash \Bigl( \Dslash[\Dslash,[\Dslash,L_f]]
+ [\Dslash,[\Dslash,L_f]]\Dslash
\\
&\hspace{12em} + 2\rho|\Dslash|\,[\Dslash,L_f]
- 2\rho[\Dslash,L_f]\,|\Dslash| \Bigr)
\frac{1}{(|\Dslash|+\rho)^2 + \la} \,\sqrt{\la}\,d\la.
\end{align*}
Since $\Dslash (|\Dslash|+\rho)^{-1}$ is bounded,
Corollary~\ref{cl:interpolation} shows that
$$
\frac{\Dslash}{|\Dslash|+\rho}\,[\Dslash,L_f]\,(1 + \Dslash^2)^{-k}
\in \L^p(\H)  \sepword{whenever} k > n/2p.
$$
For the other four summands, for example for the first one,
one gets (and similarly for the three others):
\begin{align*}
\biggl\| \frac{1}{\pi} \int_0^\infty
&\frac{\Dslash^2}{((|\Dslash|+\rho)^2 + \la)^2}\,
[\Dslash, [\Dslash, L_f]] (1 + |\Dslash|)^{-k}
\frac{1}{(|\Dslash|+\rho)^2 + \la} \,\sqrt{\la}\,d\la \biggr\|_p
\\
&\leq \bigl\| [\Dslash, [\Dslash, L_f]](1 + |\Dslash|)^{-k} \bigr\|_p
\frac{1}{\pi} \int_0^\infty
\biggl\| \frac{\Dslash^2}{(|\Dslash|+\rho)^2 + \la} \biggr\|
\biggl\|\frac{1}{(|\Dslash|+\rho)^2 + \la}\biggr\|^2
\,\sqrt{\la}\,d\la
\\
&\leq \bigl\| [\Dslash, [\Dslash, L_f]](1 + |\Dslash|)^{-k} \bigr\|_p
\frac{1}{\pi} \int_0^\infty \frac{\sqrt{\la}}{(\rho^2 + \la)^2} \,d\la
\\
&= \frac{1}{2\rho}\,
\bigl\| L_{\Dslash^2 f}\, (1 + |\Dslash|)^{-k} \bigr\|_p
\end{align*}
which is again finite, using the same Corollary.

For $[(1 + \Dslash^2)^{1/2},L_f]$, the proof goes along the same lines,
using the spectral representation \eqref{eq:spectral-sqrt} applied to
the positive operator $(1 + \Dslash^2)^{1/2}$.
\end{proof}

\subsection{Weak Schatten-class estimates}

We prove now that, as expected, noncompact isospectral deformations of
$n$-dimensional spin manifolds have spectral dimension~$n$ in the
sense of~\cite{Himalia}. The following Proposition uses the estimate
\eqref{eq:kernel2} to get an improved version of the Cwikel inequality
obtained in~\cite{Himalia}.

\begin{prop}
\label{pr:weak-schatten}
Let $f\in\Coo_c(M)$. Then 
$$
L_f\, (1 + \Delta)^{-1/2} \,L_{\bar f} \in \L^{n,\infty}(\H_r).
$$
\end{prop}
\begin{proof}
Choose a number $m$ with $0 < m < 1$. We define positive operators
\begin{align*}
A_k
&:= L_f \int_0^{m^{2k}} t^{-1/2}\, e^{-t(1+\Delta)} \,dt \ L_{\bar f},
\\
B_k
&:= L_f \int_{m^{2k}}^1 t^{-1/2}\, e^{-t(1+\Delta)} \,dt \ L_{\bar f},
\\
C &:= L_f \int_1^\infty t^{-1/2}\, e^{-t(1+\Delta)} \,dt \ L_{\bar f},
\end{align*}
for each $k \in \N$ (the most suitable value of~$k$ will be chosen
later). Their sum is
$A_k + B_k + C = \Ga(\half)\, L_f (1 + \Delta)^{-1/2} L_{\bar f}$ for
each $k \in \N$.

We note first that $C$ is in all Schatten classes $\L^p(\H_r)$ for
$p\geq 1$. Indeed, using Theorem~\ref{th:HS-norm} and
\eqref{eq:kernel2}, we get
\begin{align*}
\|C\|_1
&= \biggl\| L_f\biggl( \int_1^\infty t^{-1/2} e^{-t(1 + \Delta)} \,dt
\biggr)^{1/2} \biggr\|_2^2
= \biggl\| M_f\biggl( \int_1^\infty t^{-1/2} e^{-t(1 + \Delta)} \,dt
\biggr)^{1/2} \biggr\|_2^2
\\
&= \Tr\biggl(
M_{|f|^2} \int_1^\infty t^{-1/2} e^{-t(1 + \Delta)} \,dt \biggr)
\\
&= \int_M |f(p)|^2 \int_1^\infty t^{-1/2} e^{-t}\, K_t(p,p) 
\,\mu_g(p) \,dt
\\
&\leq c \int_M |f(p)|^2 \,\mu_g = c\, \|f\|_2^2.
\end{align*}
Thus $C \in \L^n(\H_r) \subset \L^{n,\infty}(\H_r)$.

Moreover, we can bound $A_k$ in the uniform norm:
$$
\|A_k\|
\leq \|L_f\|^2 \int_0^{m^{2k}} t^{-1/2} \|e^{-t(1 + \Delta)}\| \,dt
\leq \|L_f\|^2 \int_0^{m^{2k}} t^{-1/2} \,dt = 2\,\|L_f\|^2\,m^k.
$$
By Theorem~\ref{th:HS-norm} as above and \eqref{eq:kernel2}, we can
also estimate $B_k$ in the trace norm:
\begin{align*}
\|B_k\|_1
&= \int_{m^{2k}}^1 t^{-1/2} e^{-t}
\int_M |f(p)|^2 \,K_t(p,p) \,\mu_g(p) \,dt
\\
&\leq c \,\int_M |f(p)|^2 \,\mu_g(p)
\int_{m^{2k}}^1 t^{-(n+1)/2} \,dt
\\
&= c \, \|f\|_2^2 \, \frac{2}{n-1}\, (m^{-k(n-1)} - 1)
\\
&\leq c' \, \|f\|_2^2 \, m^{-k(n-1)},
\end{align*}
since $m < 1$.

By Fan's inequality ---see \cite[III.6.5]{Bhatia} or
\cite{SimonTrace}---, we can estimate the $j$-th singular value of
$D := A_k + B_k$:
\begin{align*}
\mu_j(D) &= \mu_j(A_k + B_k) \leq \mu_1(A_k) + \mu_j(B_k)
\\
&\leq \|A_k\| + j^{-1}\|B_k\|_1
\\
&\leq 2\,\|L_f\|^2\,m^k + c'\,\|f\|_2^2 \,\, j^{-1}\,m^{k(1-n)} .
\end{align*}
Now, given $j$ and $m < 1$, one can choose $k \in \N$ such that
$m^k \leq j^{-1/n} < m^{k-1}$. Thus 
$j^{-1}\,m^{-k(n-1)} < m^{(k-1)n}\, m^{-k(n-1)} = m^{-n}\,m^{k}$ and finally
$$
\mu_j(D) \leq c(f,n,m) \, j^{-1/n},
$$
which concludes the proof since
$L_f\,(1 + \Delta)^{-1/2}\,L_{\bar f} = \Ga(\half)^{-1} (C + D)$.
\end{proof}

This result has an immediate corollary.

\begin{cor}
\label{cl:weak-schatten}
Let $f,h \in \Coo_c(M)$. Then  
$L_f\,(1 + \Delta)^{-1/2}\,L_h \in \L^{n,\infty}(\H_r)$.
\end{cor}

\begin{proof}
Polarization: add up
$L_{(f + (-i)^k \bar h)} \, (1 + \Delta)^{-1/2}\, L_{(\bar f + i^k h)}$
for $k = 0,1,2,3$.
\end{proof}

Again, this result lifts to the Hilbert space of square-integrable
spinors.

\begin{cor}
\label{cl:weak-schatten-bis}
Both $L_f \,(1 + \Dslash^2)^{-1/2}\, L_h$ and
$L_f \,(1 + |\Dslash|)^{-1}\, L_h$ lie in $\L^{n,\infty}(\H)$
whenever $f,h \in \Coo_c(M)$.
\end{cor}

\begin{proof}
Decompose the second operator as
\begin{align*}
L_f \,(1 + |\Dslash|)^{-1}\, L_h
&= L_f(1 + \Dslash^2)^{-1/2}L_h\frac{(1 + \Dslash^2)^{1/2}}
{1 + |\Dslash|} + L_f(1 + \Dslash^2)^{-1/2}
\biggl[\frac{(1 + \Dslash^2)^{1/2}}{1 + |\Dslash|}, L_h\biggr]
\\
&= L_f (1 + \Dslash^2)^{-1/2} L_h
\frac{(1 + \Dslash^2)^{1/2}}{1 + |\Dslash|}
- L_f (1 + |\Dslash|)^{-1} [|\Dslash|,L_h] (1 + |\Dslash|)^{-1}
\\
&\qquad\qquad
+ L_f(1 + \Dslash^2)^{-1/2}\big[(1 + \Dslash^2)^{1/2},L_h\big]
(1 + |\Dslash|)^{-1}.
\end{align*}
Since $L_f(1 + |\Dslash|)^{-1}$, $[|\Dslash|,L_h](1 + |\Dslash|)^{-1}$,
$L_f(1 + \Dslash^2)^{-1/2}$ and
$[(1 + \Dslash^2)^{1/2},L_h](1 + |\Dslash|)^{-1}$ all lie in
$\L^{2n}(\H)$ by Lemma~\ref{lem:plenty}, and since
$(1 + \Dslash^2)^{1/2}(1 + |\Dslash|)^{-1}$ is bounded, it is enough to
prove the case of $L_f \,(1 + \Dslash^2)^{-1/2}\, L_h$.

Using the spectral identity \eqref{eq:spectral-sqrt} and the
Lichnerowicz formula once more, we find that
\begin{align*}
L_f\, (1 + \Dslash^2)^{-1/2} \,L_h
&= L_f\, \frac{1}{\pi} \int_0^\infty
\frac{(1 + \Dslash^2)^{-1}}{(1 + \Dslash^2)^{-1} + \la}
\,\frac{d\la}{\sqrt{\la}} \, L_h
\\
&= L_f\, \frac{1}{\pi} \int_0^\infty
\frac{(1 + \Delta)^{-1} (1 - \frac{1}{4}R (1 + \Dslash^2)^{-1})}
{(1 + \Delta)^{-1} (1 - \frac{1}{4}R (1 + \Dslash^2)^{-1}) + \la}
\,\frac{d\la}{\sqrt{\la}} \,L_h
\\
&= L_f\, \frac{1}{\pi} \int_0^\infty \biggl(
\frac{(1 + \Delta)^{-1}}{(1 + \Delta)^{-1} + \la}
+ \frac{1}{4} \frac{(1 + \Delta)^{-2}}{(1 + \Delta)^{-1} + \la} R
\frac{(1 + \Dslash^2)^{-1}}{(1 + \Dslash^2)^{-1} + \la}
\\
&\hspace{6em} - \frac{1}{4} (1 + \Delta)^{-1} R
\frac{(1 + \Dslash^2)^{-1}}{(1 + \Dslash^2)^{-1} + \la} \biggl)
\frac{d\la}{\sqrt{\la}} \,L_h
\\
&= L_f\, (1 + \Delta)^{-1/2} \,L_h
 + \frac{1}{4\pi} L_f \int_0^\infty \biggl(
\frac{(1 + \Delta)^{-2}}{(1 + \Delta)^{-1} + \la} R
\frac{(1 + \Dslash^2)^{-1}}{(1 + \Dslash^2)^{-1} + \la}
\\
&\hspace{6em} - (1 + \Delta)^{-1} R
\frac{(1 + \Dslash^2)^{-1}}{(1 + \Dslash^2)^{-1} + \la} \biggr)
\frac{d\la}{\sqrt{\la}} \,L_h.
\end{align*}
The first term lies in $\L^{n,\infty}(\H)$ by
Corollary~\ref{cl:weak-schatten} and the two others are in $\L^n(\H)$
since
\begin{align*}
\biggl\| L_f & \int_0^\infty
\frac{(1 + \Delta)^{-2}}{(1 + \Delta)^{-1} + \la} R
\frac{(1 + \Dslash^2)^{-1}}{(1 + \Dslash^2)^{-1} + \la}
\frac{d\la}{\sqrt{\la}} \,L_h \biggr\|_n
\\
&\leq \|L_f(1 + \Delta)^{-2}\|_n \|R\| \|(1 + \Dslash^2)^{-1} L_h\|
\int_0^\infty \biggl\| \frac{1}{(1 + \Delta)^{-1} + \la} \biggr\|
\biggl\| \frac{1}{(1 + \Dslash^2)^{-1} + \la} \biggr\|
\frac{d\la}{\sqrt{\la}}
\\
&\leq \|L_f(1 + \Delta)^{-2}\|_n \|R\| \|L_h\|
\int_0^\infty \frac{1}{(1 + \la)^2} \frac{d\la}{\sqrt{\la}},
\end{align*}
which is finite by Proposition~\ref{pr:interpolation}. 
Also, by Proposition~\ref{pr:interpolation} and
Corollary~\ref{cl:interpolation},
\begin{align*}
\biggl\| L_f & \int_0^\infty (1 + \Delta)^{-1} R
\frac{(1 + \Dslash^2)^{-1}}{(1 + \Dslash^2)^{-1} + \la}
\frac{d\la}{\sqrt{\la}} L_h \biggr\|_n
\\
&\leq \|R\| \|L_f\,(1 + \Delta)^{-1}\|_{2n} \,
\|(1 + \Dslash^2)^{-1}\,L_h\|_{2n}
\int_0^\infty \frac{1}{1 + \la} \,\frac{d\la}{\sqrt{\la}}
\end{align*}
is finite. Since $\L^n(\H) \subset \L^{n,\infty}(\H)$, the proof is
complete.
\end{proof}

\section{Dixmier trace computation: periodic case}
\label{sec:Trw-periodic}

In this section, we shall see that the Dixmier traces $\Trw$ ---see
\cite{Book,Polaris} for the precise definition--- give rise to an
invariant for the deformation, with exactly the same role as the
ordinary trace for the Hilbert--Schmidt-norm as seen in section
\ref{sec:Hilb-anal}. Before giving a proof of this claim, namely that
\begin{equation}
\Trw(L_f\,(1 + \Dslash^2)^{-n/2}) = \Trw(M_f\,(1 + \Dslash^2)^{-n/2}),
\sepword{for all} f \in \Coo_c(M),
\label{eq:Dixtr-invt}
\end{equation}
(or at the scalar level, i.e., when $L_f$ is acting on $\H_r$, with
$(1 + \Delta)^{-n/2}$ replaced by $(1 + \Dslash^2)^{-n/2}$), we give
an heuristic argument to see why this result is plausible. To this
end, we will take advantage of the possibility of viewing $L_f$, for
$f \in \Coo_c(M)$, as an integral of bounded operators given by
\eqref{eq:int-pres}. Using this presentation for $L_f$, the trace
property of the Dixmier trace and the commutativity of the Dirac
operator (or the Laplacian) with the unitaries $V_z$ (or
$V_{\tilde z}$), the result would be straightforward if we could swap
the Dixmier trace with the Lebesgue integral:
\begin{align*}
\Trw((L_f\,(1 + \Dslash^2)^{-n/2})
&= (2\pi)^{-l}\Trw \biggl( \int_{\R^{2l}} e^{-iy z}\,
 V_{\half\Th y}\, M_f \,V_{-\half\Th y-z}
\,d^ly \,d^lz\,(1 + \Dslash^2)^{-n/2} \biggr)
\\
&= (2\pi)^{-l} \int_{\R^{2l}} e^{-iy z} \Trw(
V_{\half\Th y}  M_f \,(1 + \Dslash^2)^{-n/2}\, V_{-\half\Th y-z})
\,d^ly \,d^lz
\\
&= (2\pi)^{-l} \int_{\R^{2l}} e^{-iy z} \Trw(
M_f \,(1 + \Dslash^2)^{-n/2}\, V_{-z}) \,d^ly \,d^lz
\\
&= \Trw \biggl( M_f \,(1 + \Dslash^2)^{-n/2} \int_{\R^l}
\delta_0(z)\, V_{-z} \,d^lz \biggr)
\\
&= \Trw\big( M_f\,(1 + \Dslash^2)^{-n/2}\big).
\end{align*}

However, this exchange of the Dixmier trace with the integral is not
rigorous, since the integrals are oscillatory and Dixmier traces do
not in general obey dominated convergence.

For the ordinary trace, the situation is better since such an exchange
can be justified by using a family of strongly convergent
regularizers. For example, one can use $\{M_{u_k}\}$, where
$\{u_k\}_{k\in\N}$ is an approximate unit for $\Coo_c(M)$, that
is, an increasing family of nonnegative compactly supported functions
such that $u_k \uparrow 1$ pointwise on $M$, so that
$\slim M_{u_k} = 1$. Then, the integrals in the product
$M_{u_k} L_f (1 + \Dslash^2)^{-k} L_{\bar f} M_{u_k}$ (with $k > n/2$)
become Bochner integrals for the trace-norm, with uniform bound on its
trace-norm. Finally, by \cite[Prop.~2]{DeiftS}, we obtain that the
strong limit
$$
\slim M_{u_k} L_f (1 + \Dslash^2)^{-k} L_{\bar f} M_{u_k} =
L_f(1 + \Dslash^2)^{-k}L_{\bar f}
$$
is trace-class as well, with the same trace-norm bound as for the
family $M_{u_k} L_f (1 + \Dslash^2)^{-k} L_{\bar f} M_{u_k}$. This
gives another proof of Theorem~\ref{th:HS-norm}.

Such an approach fails for the Dixmier trace, since these natural
regularizers give rise to trace-class operators in some cases. This is
for instance the case for Moyal planes, since one can prove that
$L_f\,(1 + \Dslash^2)^{-k}\,M_{u_k}$ is trace-class for all $k \geq 0$
whenever $f,u_k \in \SS(\R^l)$, so they have vanishing Dixmier trace.

In the aperiodic case, we shall prove this condition indirectly, using
a zeta-residue argument to evaluate the left hand side
of~\eqref{eq:Dixtr-invt} as the same ordinary integral which is known
to give the value of the right hand side~\cite[Prop.~15]{RennieSumm}.
Before that, we first establish the result in the easier periodic
case, for which no hard analysis is needed. The spectral subspace
decomposition of $f \in \Coo_c(M)$ gives a direct access to the
Dixmier traceability of the operators $L_f\,(1 + \Delta)^{-n/2}$
acting on $\H_r = L^2(M,\mu_g)$ and $L_f\,(1 + \Dslash^2)^{-n/2}$
acting on $\H = L^2(M,S)$, as well as to the value of their Dixmier
traces.

\begin{prop}
\label{pr:periodic}
Let $\a$ be an effective isometric smooth action of $\T^l$ on $M$,
with $l \geq 2$, and let $f \in \Coo_c(M)$. Then the operator
$L_f\,(1 + \Delta)^{-n/2}$ is Dixmier traceable on $\H_r$, and the
value of its Dixmier trace is independent of $\omega$:
$$
\Trw(L_f\,(1 + \Delta)^{-n/2})
= C'(n)\,\delta_{0,r}\,\int_Mf_r \, \mu_g
= C'(n)\,\int_M f_0 \, \mu_g,
$$
where $C'(n) := \Omega_n/n\,(2\pi)^n$, $\Omega_n$ is the volume of the
unit sphere in $\R^n$, and $f = \sum_r f_r$ is the decomposition
\eqref{eq:Peter-Weyl} of $f$ in homogeneous components.
\end{prop}

\begin{proof}
Each $f_r$ satisfies $\a_z(f_r) = e^{-iz r}\,f_r$ for all
$z \in \T^l$. Since $[M_{f_r}, V_z] = M_{f_r} (1 - e^{-iz r}) V_z$,
we see that $[M_{f_r}, V_{-\half\Th r}] = 0$ by skew-symmetry of the
deformation matrix.

By \cite[Prop.~15]{RennieSumm}, $M_f\,(1 + \Delta)^{-n/2}$ lies in
$\L^{1,\infty}(\H_r)$, and moreover
$$
\|M_f\,(1 + \Delta)^{-n/2}\|_{1,\infty} \leq C_1(n)\, \|f\|_\infty.
$$
This estimate is obtained by a (finite) partition of unity on the
compact set $(\supp f)$ on applying Weyl's theorem. We thus obtain,
using \eqref{eq:Lf-decomp},
\begin{align*}
\|L_f\,(1 + \Delta)^{-n/2}\|_{1,\infty}
&\leq \sum_{r\in\Z^l}
\|M_{f_r} V_{-\half\Th r}\,(1 + \Delta)^{-n/2}\|_{1,\infty}
\\
&\leq \sum_{r\in\Z^l} \|M_{f_r}\,(1 + \Delta)^{-n/2}\|_{1,\infty}
\\
&\leq C_1(n) \sum_{r\in\Z^l} \|f_r\|_\infty,
\end{align*}
since each $f_r$ is compactly supported with support contained in
$\T^l \cdot (\supp f)$. Those estimates give the Dixmier traceability,
since the spectral-subspace decomposition is
$\|\cdot\|_\infty$-convergent.

To compute the Dixmier trace, it remains to remark that for all
$z \in \T^l$,
\begin{align*}
\Trw(L_{f_r}\,(1 + \Delta)^{-n/2})
&=\Trw(V_z\,M_{f_r}\,V_{-\half\Th r}\,(1+\Delta)^{-n/2}\,V_{-z})
\\
&= \Trw(M_{\a_z(f_r)}\,V_{-\half\Th r}\,(1 + \Delta)^{-n/2})
\\
&= e^{-iz r} \Trw(M_{f_r}\,V_{-\half\Th r}\,(1+\Delta)^{-n/2}),
\end{align*}
and therefore it must vanish unless $r = 0$ because of
\eqref{eq:Lf-decomp}. Thus,
$$
\Trw\big(L_{f_r}\,(1 + \Delta)^{-n/2}\big)
= \Trw\big(M_{f_0}\,(1 + \Delta)^{-n/2}\big)\,\delta_{0,r}
= C'(n)\,\delta_{0,r}\,\int_M f_0 \, \mu_g.
$$
The last equality is obtained, as in \cite[Prop.~15]{RennieSumm}, by
computation of the Wodzicki residue of the operator
$M_f(1 + \Delta)^{-n/2}$.
\end{proof}

\begin{cor}
\label{cl:periodic}
Under the same hypothesis, the operator
$L_f\,(1 + \Dslash^2)^{-n/2}$ is Dixmier traceable on $\H$ for
$f \in \Coo_c(M)$; furthermore, the value of its Dixmier trace is
independent of $\omega$:
$$
\Trw(L_f\,(1 + \Dslash^2)^{-n/2}) =
C(n)\,\delta_{0,r}\,\int_M f_r \,\mu_g  = C(n)\,\int_M f_0 \,\mu_g,
$$
where $C(n) := 2^{\piso{n/2}} \Omega_n/n\,(2\pi)^n$, with
$2^{\piso{n/2}}$ being the rank of the spinor bundle.
\end{cor}

\begin{proof}
Using the Lichnerowicz formula $\Dslash^2 = \Delta^S + \tfrac{1}{4}R$,
the Dixmier traceability is obtained by comparison:
\begin{equation}
\label{lich}
(1 + \Dslash^2)^{-1} = (1 + \Delta^S)^{-1}
\bigl(1 - \tfrac{1}{4} R\,(1 + \Dslash^2)^{-1} \bigr).
\end{equation}
For the computation of the Dixmier trace, one can apply previous
arguments. We obtain the result, using that, modulo the factor
$2^{\piso{n/2}}$, the principal symbols of $(1 + \Dslash^2)^{-n/2}$
and $(1 + \Delta)^{-n/2}$ are the same as seen in \eqref{lich}. Thus,
the operators $M_{f_r}(1 + \Dslash^2)^{-n/2}$ and
$M_{f_r}(1 + \Delta)^{-n/2}$ have the same Wodzicki residue, up to
that constant factor.
\end{proof}

\section{Dixmier trace computation: aperiodic case}
\label{sec:Trw-aperiodic}

This subsection is devoted to the proof of the following theorem.

\begin{thm}
\label{th:Dix-tr}
Let $M$ be a noncompact, connected, complete Riemannian spin manifold
satisfying Condition~\ref{cn:heat-ker}, with bounded scalar curvature.
Suppose further that $M$ is endowed with a smooth isometric and proper
action of~$\R^l$. If $f \in \Coo_c(M)$, then
$L_f\,(1 + |\Dslash|)^{-n}$ lies in $\L^{1,\infty}(\H)$ and is a
measurable operator; the common value of its Dixmier traces is
$$
\Trw \bigl( L_f\, (1 + |\Dslash|)^{-n} \bigr)
= C(n) \int_M f(p) \,\mu_g(p),
$$
where $C(n) = 2^{\piso{n/2}}\,\Omega_n/n\,(2\pi)^n$.
\end{thm}

In the aperiodic case, the manifold $M$ is necessarily of the form
$V \x \R^l$, where the group $\R^l$ acts by translation on the second
direct factor. Indeed, proper actions of the additive group $\R^l$ are
automatically free, because the only compact subgroup of $\R^l$ is the
trivial subgroup $\{0\}$. Thus, the projection on the orbit space
$\pi: M \to V := M/\R^l$ defines a principal $\R^l$-bundle projection
\cite[Thm.~1.11.4]{DuistermaatK}. We remark that properness of the
action was crucially used in Proposition~\ref{pr:kernel} to show that
twisted multiplication operators are bounded. But a principal
$\R^l$-bundle has a smooth global section and so it is automatically
trivializable: see \cite[16.14.5]{DieudonneIII}, for instance.

Thus we write $M = V \x \R^l$, where $V$ is a smooth (not
necessarily compact) manifold of dimension $k = n - l$, which carries
a Riemannian metric, induced from that of $M$, and $\pi: M \to V$
is just the projection on the first factor. If 
$\{\phi_j\}_{j\in J}$ is any locally finite partition of unity on~$V$ 
consisting of smooth compactly supported functions, then by setting
$\psi_j := \phi_j \circ \pi$, we obtain an $\a$-invariant partition of
unity $\{\psi_j\}$ on $M$. For any $f \in \Coo_c(M)$, the sum
$f = \sum_j f \psi_j$ is finite because $\supp f$ is compact; since
each $\psi_j$ is $\a$-invariant, we directly obtain
$$
L_f = \sum_j L_{f\psi_j} = \sum_j L_f\,M_{\psi_j}.
$$
Thus, when dealing with operators of the form $L_f\,h(\Dslash)$, we
lose no generality by restricting to a single coordinate chart of~$V$;
so we shall assume from now on that $V$ is an open ball in~$\R^k$.

We denote by $\hat x := (x^1,\dots,x^k) \in V$ and
$\bar x := (x^{k+1},\dots,x^n) \in \R^l$ respectively the transverse
and longitudinal local coordinates on~$M$. It is immediate that the
operator $L_f$ is pseudodifferential, with symbol
\begin{equation}
\sigma[L_f](\hat x,\bar x;\hat\xi,\bar\xi)
= f(\hat x,\bar x - \half\Th\bar\xi).
\end{equation}
Indeed, for any vector $\psi \in \H$, Definition~\ref{df:left-multn}
shows that
\begin{align*}
L_f\psi(\hat x,\bar x)
&= (f \Mop \psi)(\hat x,\bar x)
= (2\pi)^{-l} \int_{\R^{2l}} e^{-i\bar\xi \bar y}
\a_{\half\Th\bar\xi}(f)(\hat x,\bar x)\,
V_{-\bar y}\psi(\hat x,\bar x) \,d^l\bar\xi \,d^l\bar y
\\
&= (2\pi)^{-l} \int_{\R^{2l}} e^{-i\bar\xi \bar y}
f(\hat x,\bar x - \half\Th\bar\xi)\,
\psi(\hat x,\bar x + \bar y) \,d^l\bar\xi \,d^l\bar y
\\
&= (2\pi)^{-n} \int_{\R^{2n}}
e^{-i\bar\xi(\bar y-\bar x)} e^{-i\hat\xi(\hat y-\hat x)} \,
f(\hat x,\bar x - \half\Th\bar\xi)\, \psi(\hat y,\bar y)
\,d^l\bar\xi \,d^l\bar y \,d^k\hat\xi \,d^k\hat y.
\end{align*}

\begin{prop}
\label{pr:matrix-basis}
Under the hypotheses of Theorem~\ref{th:Dix-tr}, if $f \in \Coo_c(M)$
then $L_f(1 + |\Dslash|)^{-n}$ lies in $\L^{1,\infty}(\H)$.
\end{prop}

\begin{proof}
For fixed $\hat x$, the function $\bar x \mapsto f(\hat x,\bar x)$
lies in $\Coo_c(\R^l)$, so it can be decomposed in the Wigner
eigentransition basis $\{f_{mn}\}$, indexed by $m,n \in \N^{l/2}$ (see
\cite{Himalia,Phobos,Deimos} and recall that $l$ is even):
$$
f(\hat x,\bar x) = \sum_{m,n} c_{mn}(\hat x) \,f_{mn}(\bar x),
$$
where the matrix coefficients $c_{mn}$ lie in $\Coo_c(V)$.

Given two functions
$f(\hat x,\bar x) = \sum c_{mn}(\hat x) \,f_{mn}(\bar x)$,
$h(\hat x,\bar x) = \sum d_{mn}(\hat x) \,f_{mn}(\bar x)$ of this
form, their twisted product may thus be expressed as a matrix product
in the $\bar x$~variables:
\begin{equation}
(f \Mop h)(\hat x,\bar x)
= \sum_{m,n,k} c_{mk}(\hat x)\,d_{kn}(\hat x) \,f_{mn}(\bar x).
\label{eq:Moyal-matricial}
\end{equation}
The operator $L_f$ can then be viewed as an element of the algebra
$M_\infty(\Coo(V))$ with rapidly decreasing $\Coo(V)$-valued matrix
elements.

Thus, one can extend the strong factorization property \cite{Phobos}
of the algebra $(\SS(\R^l),\Mop)$ to this context: for all
$f \in \Coo_c(M)$, there exist $h, k \in \Coo(M)$ which are Schwartz
functions in the $\bar x$~variables, such that
\begin{equation}
f(\hat x,\bar x) = (h \Mop k)(\hat x,\bar x).
\label{eq:strong-factorn}
\end{equation}
By iterated factorization, allowing to write $f$ as a product of $n$
such functions, and by taking iterated commutators, exactly as in
Corollary 4.12 and Lemma 4.13 of~\cite{Himalia}, we can express each
$L_f(1 + |\Dslash|)^{-n}$ as a product of $n$ terms of the form
$L_h (1 + |\Dslash|)^{-1} L_k$, each lying in $\L^{n,\infty}(\H)$ by
Corollary~\ref{cl:weak-schatten-bis}, plus an extra term in
$\L^1(\H)$. Finally, by the H\"older inequality for weak Schatten
classes, we conclude that
$L_f(1 + |\Dslash|)^{-n} \in \L^{1,\infty}(\H)$.
\end{proof}

We may also introduce a system of local units \cite{RennieSmooth} for
the twisted product by a straightforward extension of a construction
in~\cite{Himalia}.

\begin{defn}
The manifold $V$ may be expressed as a union of compact subsets $C_i$
with each contained in the interior of $C_{i+1}$; define $\chi_i := 1$
on~$C_i$ and $\chi_i := 0$ elsewhere. For each $K \in \N$, define a
function $e_K$ on~$M$ by
$$
e_K(\hat x,\bar x)
:= \sum_{|n|\leq K} \chi_K(\hat x) \,f_{nn}(\bar x),
$$
where $|n| = n_1 +\cdots+ n_{l/2}$. Then $e_K$ is real-valued and
$e_K \Mop e_K = e_K$ by using \eqref{eq:Moyal-matricial} to compute
the twisted product, and $L_{e_K}$ is defined as an orthogonal
projector on~$\H$. Next, let $f_K := e_K \Mop f \Mop e_K$, or more
explicitly,
\begin{equation}
f_K(\hat x,\bar x)
:= \sum_{|m|,|n|\leq K} c_{mn}(\hat x) \,f_{mn}(\bar x).
\label{eq:f-trunc}
\end{equation}
By construction, $e_K \Mop f_K = f_K \Mop e_K = f_K$.
\end{defn}

The operator $L_{e_K} (1 + |\Dslash|)^{-n} L_{e_K}$ is Dixmier
traceable: in Proposition~\ref{pr:weak-schatten} and subsequently, one
can replace $f$ by $e_K$ even though the latter is not in $\Coo_c(M)$,
since its square-integrability is guaranteed at each step, and the
factorization argument following \eqref{eq:strong-factorn} goes
through because $e_K$ is idempotent. The trace property of the Dixmier
trace now yields
$$
\Trw\bigl( L_{f_K} (1 + |\Dslash|)^{-n} \bigr) =
\Trw\bigl( L_{f_K} L_{e_K} (1 + |\Dslash|)^{-n} L_{e_K} \bigr).
$$
Since $L_{f_K}$ is bounded, Theorem~5.6 of \cite{CareyPS} shows that
if the following limit exists:
$$
\lim_{s\downarrow 1} (s - 1)
\Tr\bigl(L_{f_K}(L_{e_K} (1 + |\Dslash|)^{-n} L_{e_K})^s \bigr),
$$
then it will coincide with the value of any Dixmier trace of
$L_{f_K} (1 + |\Dslash|)^{-n}$.

\begin{lem}
\label{lm:zeta-control}
The trace norm
\begin{equation}
\bigl\| L_{f_K} (L_{e_K} (1 + |\Dslash|)^{-n} L_{e_K})^s
- L_{f_K} (1 + |\Dslash|)^{-ns} \bigl\|_1
\label{eq:zeta-control}
\end{equation}
is a bounded function of~$s$, for $1 \leq s \leq 2$.
\end{lem}

\begin{proof}
Write $s =: 1 + \eps$, with $0 < \eps \leq 1$. We use the following
spectral representation, generalizing \eqref{eq:spectral-sqrt}, for
fractional powers of a positive operator~$A$:
$$
A^\eps = \frac{\sin\pi\eps}{\pi}
\int_0^\infty A \, (1 + \la A)^{-1} \, \la^{-\eps} \,d\la.
$$
Since $L_{e_K}$ is an orthogonal projector and
$L_{f_K} L_{e_K} = L_{f_K}$, we can write
\begin{align*}
L_{f_K} (L_{e_K} (1 + |\Dslash|)^{-n} L_{e_K})^s
&= L_{f_K} L_{e_K} (1 + |\Dslash|)^{-n} L_{e_K} \,
(L_{e_K} (1 + |\Dslash|)^{-n} L_{e_K}\big)^\eps
\\
&= L_{f_K} (1 + |\Dslash|)^{-n}
(L_{e_K} (1 + |\Dslash|)^{-n} L_{e_K})^\eps.
\end{align*}
Hence,
\begin{align}
L_{f_K} & (L_{e_K} (1 + |\Dslash|)^{-n} L_{e_K})^s
- L_{f_K} (1 + |\Dslash|)^{-ns}
\label{eq:power} \\
&=
L_{f_K} (1 + |\Dslash|)^{-n}\, \frac{\sin\pi\eps}{\pi}
\int_0^\infty \biggl(
\frac{L_{e_K} (1 + |\Dslash|)^{-n} L_{e_K}}
     {1 + \la L_{e_K} (1 + |\Dslash|)^{-n} L_{e_K}}
- \frac{(1 + |\Dslash|)^{-n}}{1 + \la(1 + |\Dslash|)^{-n}} \biggr)
\,\la^{-\eps} \,d\la.
\nonumber
\end{align}
The first fraction in parenthesis may be rewritten as
$$
\bigl( (1 + |\Dslash|)^n + \la T_K \bigr)^{-1} \,T_K,
$$
where
$$
T_K := (1 + |\Dslash|)^n L_{e_K} (1 + |\Dslash|)^{-n} L_{e_K}.
$$
Since $L_{e_K}$ is a projector, we get
\begin{align}
T_K &= L_{e_K}^2
+ [(1 + |\Dslash|)^n, L_{e_K}]\, (1 + |\Dslash|)^{-n} L_{e_K}
\nonumber \\
&= L_{e_K} + \sum_{0\leq k<r\leq n}  \binom{n}{r}
|\Dslash|^k \,[|\Dslash|, L_{e_K}] \,|\Dslash|^{r-k-1}
(1 + |\Dslash|)^{-n} \,L_{e_K}
\nonumber \\
&=: L_{e_K} + \sum_{0\leq k<r\leq n} A_{rk}.
\label{eq:teux}
\end{align}
The crucial issue in showing the difference \eqref{eq:power} to be
$\eps$-uniformly trace-class is that, excepting the first summand
in~\eqref{eq:teux} which is merely bounded, all the other summands
$A_{rk}$ are compact. More precisely, using
Proposition~\ref{pr:interpolation} (plus routine commutations), we
can check that each $A_{rk} \in \L^p(\H)$ for all $p > n$.

Following the procedure of Rennie \cite[Thm.~12]{RennieSumm}, we
reduce the difference of fractions in \eqref{eq:power} as follows:
\begin{align*}
& \frac{L_{e_K} (1 + |\Dslash|)^{-n} L_{e_K}}
     {1 + \la L_{e_K} (1 + |\Dslash|)^{-n} L_{e_K}}
- \frac{(1 + |\Dslash|)^{-n}}{1 + \la(1 + |\Dslash|)^{-n}}
\\
&= ((1 + |\Dslash|)^n + \la T_K)^{-1} \,T_K
- ((1 + |\Dslash|)^n + \la)^{-1}
\\
&= \bigl( ((1 + |\Dslash|)^n + \la T_K)^{-1}
   - ((1 + |\Dslash|)^n + \la)^{-1} \bigr) \,T_K
+ \bigl( (1 + |\Dslash|)^n + \la \bigr)^{-1} \,(T_K - 1)
\\
&= ((1 + |\Dslash|)^n + \la)^{-1} (\la - \la T_K)
((1 + |\Dslash|)^n + \la T_K)^{-1} \,T_K
+ \bigl( (1 + |\Dslash|)^n + \la \bigr)^{-1} \,(T_K - 1)
\\
&= ((1 + |\Dslash|)^n + \la)^{-1}  \,(T_K - 1)\,
\bigl(1 - ((1 + |\Dslash|)^n + \la T_K)^{-1} \,\la T_K \bigr)
\\
&= ((1 + |\Dslash|)^n + \la)^{-1}  \,(T_K - 1)\,
\bigl(1 + \la L_{e_K} (1 + |\Dslash|)^{-n} L_{e_K} \bigr)^{-1} \,.
\end{align*}
Thus, we obtain
\begin{align*}
L_{f_K} & (L_{e_K} (1 + |\Dslash|)^{-n} L_{e_K})^s
- L_{f_K} (1 + |\Dslash|)^{-ns}
\\
&= L_{f_K} (1 + |\Dslash|)^{-n} \frac{\sin\pi\eps}{\pi}
\int_0^\infty \frac{1}{(1 + |\Dslash|)^n + \la} (T_K - 1)
\frac{1}{1 + \la L_{e_K} (1 + |\Dslash|)^{-n} L_{e_K}}
\,\la^{-\eps} \,d\la
\\
&= L_{f_K} (1 + |\Dslash|)^{-n} \frac{\sin\pi\eps}{\pi}
\int_0^\infty \frac{1}{(1 + |\Dslash|)^n + \la} \,L_{e_K} (T_K - 1)
\frac{1}{1 + \la L_{e_K} (1 + |\Dslash|)^{-n} L_{e_K}}
\,\la^{-\eps} \,d\la
\\
&\qquad + L_{f_K} \frac{\sin\pi\eps}{\pi} \int_0^\infty
\biggl[ L_{e_K},
\frac{(1 + |\Dslash|)^{-n}}{(1 + |\Dslash|)^n + \la} \biggr]
(T_K - 1) \frac{1}{1 + \la L_{e_K} (1 + |\Dslash|)^{-n} L_{e_K}}
\,\la^{-\eps} \,d\la.
\end{align*}

We now show that the second term on the right hand side is uniformly
bounded in trace norm. We write
\begin{align*}
\bigl[L_{e_K},
& (1 + |\Dslash|)^{-n} ((1 + |\Dslash|)^n + \la)^{-1} \bigr]
\\
&= [L_{e_K}, (1 + |\Dslash|)^{-n}]\, ((1 + |\Dslash|)^n + \la)^{-1}
+ (1 + |\Dslash|)^{-n} \,[L_{e_K}, ((1 + |\Dslash|)^n + \la)^{-1}],
\end{align*}
and the first of these summands yields the trace-norm estimate:
\begin{align*}
& \biggl\| L_{f_K} [L_{e_K}, (1 + |\Dslash|)^{-n}]\,
\frac{\sin\pi\eps}{\pi} \int_0^\infty \!
\frac{1}{(1+|\Dslash|)^n + \la} L_{e_K} (T_K - 1)
\frac{\la^{-\eps}}{1 + \la L_{e_K} (1 + |\Dslash|)^{-n} L_{e_K}}
\,d\la \biggr\|_1
\\
&\leq \bigl\| L_{f_K} [L_{e_K}, (1 + |\Dslash|)^{-n}] \bigr\|_1
\\
&\qquad \x \frac{\sin\pi\eps}{\pi}
\int_0^\infty \|((1+|\Dslash|)^n + \la)^{-1}\|\, \|L_{e_K}(T_K-1)\|
\bigl\| (1 + \la L_{e_K} (1+|\Dslash|)^{-n} L_{e_K})^{-1} \bigr\|
\,\la^{-\eps} \,d\la
\\
&\leq \|L_{e_K} (T_K - 1)\|\,
\bigl\| L_{f_K} [L_{e_K}, (1 + |\Dslash|)^{-n}] \bigr\|_1
\frac{\sin\pi\eps}{\pi} \int_0^\infty \frac{\la^{-\eps}}{1+\la} \,d\la
\\
&= \|L_{e_K} (T_K - 1)\|\,
\bigl\| L_{f_K} [L_{e_K}, (1 + |\Dslash|)^{-n}] \bigr\|_1 =: C_1.
\end{align*}
This constant $C_1$ is finite (and independent of $\eps$) since
\begin{equation}
L_{f_K} \,[L_{e_K}, (1 + |\Dslash|)^{-n}]
= L_{f_K} \sum_{0\leq k<r\leq n} \binom{n}{r}
\frac{|\Dslash|^k}{(1 + |\Dslash|)^n} \,[|\Dslash|, L_{e_K}]\,
\frac{|\Dslash|^{r-k-1}}{(1 + |\Dslash|)^n},
\label{eq:comm-expan}
\end{equation}
and each term of the sum lies in $\L^1(\H)$, using
Proposition~\ref{pr:interpolation} and the H\"older inequality.
Analogously, one can show that the trace-norm of
$$
L_{f_K} (1 + |\Dslash|)^{-n} \,\frac{\sin\pi\eps}{\pi} \int_0^\infty
\! [L_{e_K}, ((1 + |\Dslash|)^n + \la)^{-1}]\, L_{e_K} (T_K - 1)
\frac{\la^{-\eps}}{1 + \la L_{e_K} (1 + |\Dslash|)^{-n} L_{e_K}}
\,d\la
$$
is bounded by the constant $C_2 := \|L_{f_K} (1 + |\Dslash|)^{-n}\|
\, \|[L_{e_K}, (1 + |\Dslash|)^{-n}]\|_1$, independent of~$\eps$.

Using the expansion \eqref{eq:teux} of $T_K$,
we finally obtain
\begin{align*}
& \bigl\| L_{f_K} (L_{e_K} (1 + |\Dslash|)^{-n} L_{e_K})^s
- L_{f_K} (1 + |\Dslash|)^{-ns} \bigr\|_1
\\
&\leq \sum_{0\leq k<r\leq n} \biggl\| L_{f_K} (1 + |\Dslash|)^{-n}
\frac{\sin\pi\eps}{\pi} \int_0^\infty \frac{1}{(1+|\Dslash|)^n + \la}
L_{e_K} A_{rk}
\frac{\la^{-\eps}}{1 + \la L_{e_K} (1 + |\Dslash|)^{-n} L_{e_K}}
\,d\la \biggr\|_1
\\
&\hspace{6em} + C_1 + C_2
\\
&\leq \sum_{0\leq k<r\leq n}
\bigl\| L_{f_K} (1 + |\Dslash|)^{-n} \bigr\|_{p/(p-1)}
\, \|L_{e_K} A_{rk}\|_p \frac{\sin\pi\eps}{\pi} \int_0^\infty
\frac{\la^{-\eps}}{1 + \la} \,d\la \, + C_1 + C_2
\\
&= \sum_{0\leq k<r\leq n}
\bigl\| L_{f_K} (1 + |\Dslash|)^{-n} \bigr\|_{p/(p-1)}
\, \|L_{e_K} A_{rk}\|_p + C_1 + C_2,
\end{align*}
which is finite for $p > n$.
\end{proof}

\begin{proof}[Proof of Theorem~\ref{th:Dix-tr}]
For $1 < s \leq 2$, the operator $L_{f_K}(1 + |\Dslash|)^{-ns}$
appearing in~\eqref{eq:zeta-control} is trace-class, since it equals
the product of
$L_{f_K}(1 + |\Dslash|)^{-n} \in \L^{1,\infty}(\H)$ by
$L_{e_K} (1 + |\Dslash|)^{-n(s-1)} \in \L^p(\H)$ for
$p > 1/(s - 1)$, plus a commutator of trace class. The difference of
traces
$$
\Tr\bigl( L_{f_K} (L_{e_K} (1 + |\Dslash|)^{-n} L_{e_K})^s \bigr)
- \Tr\bigl( L_{f_K} (1 + |\Dslash|)^{-ns} \bigr)
$$
is therefore a bounded function of $s$, for $1 \leq s \leq 2$. Thus,
\begin{equation}
\lim_{s\downarrow 1} (s - 1)
\Tr\bigl( L_{f_K} (L_{e_K} (1 + |\Dslash|)^{-n} L_{e_K})^s \bigr)
= \lim_{s\downarrow 1} (s - 1)
\Tr\bigl( L_{f_K} (1 + |\Dslash|)^{-ns} \bigr).
\label{eq:zeta-res}
\end{equation}

Moreover,
\begin{equation}
\lim_{s\downarrow 1} (s - 1)
\Tr\bigl( L_{f_K} (1 + |\Dslash|)^{-ns} \bigr)
= \lim_{s\downarrow 1} (s - 1)
\Tr\bigl( L_{f_K} (1 + \Dslash^2)^{-ns/2} \bigr).
\label{eq:zeta-res2}
\end{equation}
Indeed, for $1 \leq s \leq 2$, the following operator inequalities
hold:
\begin{align*}
0 \leq (1 + \Dslash^2)^{-ns/2} - (1 + |\Dslash|)^{-ns} 
&= (1 + |\Dslash|)^{-ns} \biggl(
\Bigl( 1 + \frac{2|\Dslash|}{1+\Dslash^2} \Bigr)^{ns/2} - 1 \biggr)
\\
&\leq (1 + |\Dslash|)^{-n} \biggl(
\Bigl( 1 + \frac{2|\Dslash|}{1+\Dslash^2} \Bigr)^n - 1 \biggr)
\\
&= (1 + |\Dslash|)^{-n} \sum_{k=1}^n \binom{n}{k}
\biggl( \frac{2|\Dslash|}{1+\Dslash^2} \biggr)^k,
\end{align*}
and thus
\begin{align*}
\bigl| \Tr &\bigl( L_{f_K}\,\bigl( (1 + \Dslash^2)^{-ns/2}
- (1 + |\Dslash|)^{-ns} \bigr) \bigr) \bigr|
\\
&= \bigl| \Tr \bigl( L_{f_K}\,L_{e_K}\,\bigl( (1 + \Dslash^2)^{-ns/2}
- (1 + |\Dslash|)^{-ns} \bigr) \,L_{e_K} \bigr) \bigr|
\\
&\leq \|L_{f_K}\| \Tr\bigl( L_{e_K}\,\bigl( (1 + \Dslash^2)^{-ns/2}
- (1 + |\Dslash|)^{-ns} \bigr) \,L_{e_K} \bigr)
\\
&\leq \|L_{f_K}\| \sum_{k=1}^n \binom{n}{k} \Tr\biggl(
L_{e_K} \,(1 + |\Dslash|)^{-n} \Bigl(
\frac{2|\Dslash|}{1+\Dslash^2} \Bigr)^k \,L_{e_K} \biggr)
\\
&\leq \|L_{f_K}\| \sum_{k=1}^n \binom{n}{k}
\bigl\| L_{e_K}\,(1 + |\Dslash|)^{-n} \bigr\|_{p/(p-1)}
\biggl\| \Bigl( \frac{2|\Dslash|}{1+\Dslash^2} \Bigr)^k\,
L_{e_K} \biggr\|_p,
\end{align*}
which is finite for $p > n$.

Note that $M_{f_K} (1 + \Dslash^2)^{-ns/2}$ is also trace-class for
$s > 1$, on account of the form \eqref{eq:f-trunc} of $f_K$ on
$M = V \x \R^l$. Corollary~\ref{cr:tr-invt} now implies that
$$
\Tr\bigl( L_{f_K} (1 + \Dslash^2)^{-ns/2} \bigr)
= \Tr\bigl( M_{f_K} (1 + \Dslash^2)^{-ns/2} \bigr).
$$
The evaluation of the right hand side of \eqref{eq:zeta-res} is
therefore given by
\begin{equation}
\lim_{s\downarrow 1} (s - 1)\,
\Tr\bigl( L_{f_K}\, (1 + |\Dslash|)^{-ns} \bigr)
=\lim_{s\downarrow 1} (s - 1)\,
\Tr\bigl( M_{f_K}\, (1 + \Dslash^2)^{-ns/2} \bigr).
\label{eq:res-value}
\end{equation}
The right hand side may be rewritten as
$$
\lim_{s\downarrow 1} \frac{(s - 1)}{\Ga(\frac{ns}{2})}\,
\int_M f_K(p) \int_0^\infty t^{ns/2-1} \,e^{-t}\,
K_{e^{-t\Dslash^2}}(p,p) \,dt \,\mu_g(p).
$$
Now $\Dslash^2$ is a second-order differential operator of Laplace
type by the Lichnerowicz formula~\cite{Gilkey}, and thus
$K_{e^{-t\Dslash^2}}(p,p) =
2^{\piso{n/2}} (4\pi t)^{-n/2} + O(t^{-n/2+1})$ when $t \to 0$; the
$t$-integral from $\eps$ to $\infty$, for any $\eps < 1$, gives no
contribution thanks to the factor $e^{-t}$. (See, for instance,
\cite[Lemma~4.1.4]{Gilkey}, noting that this estimate for on-diagonal
values of the heat kernel does not depend on compactness of the
manifold.) Therefore,
\begin{align*}
\lim_{s\downarrow 1} (s - 1)
&\Tr\bigl( M_{f_K}\, (1 + \Dslash^2)^{-ns/2} \bigr)
\\
&= \lim_{s\downarrow 1} (s - 1)\,
\frac{2^{\piso{n/2}}}{(4\pi)^{n/2}\Ga(\frac{n}{2})}\,
\int_0^\infty t^{n(s-1)/2-1} \,e^{-t} \,dt \int_M f_K(p) \,\mu_g(p)
\\
&= C(n) \int_M f_K(p) \,\mu_g(p).
\end{align*}
The proportionality factor
$C(n) = 2^{\piso{n/2}}/(4\pi)^{n/2}\,\Ga(\frac{n}{2} + 1)$ is the 
same as that of Corollary~\ref{cl:periodic}.

It remains to remove the truncation induced by the projectors
$L_{e_K}$. Notice first that
$$
\Trw\bigl( (L_f - L_{f_K})\, (1 + |\Dslash|)^{-n} \bigr)
= \Trw\bigl( (1 - L_{e_K}) L_f\, (1 + |\Dslash|)^{-n} \bigr),
$$
since $L_f[L_{e_K}, (1 + |\Dslash|)^{-n}]$ is trace-class, as is
seen on replacing $f_K$ by $f$ in~\eqref{eq:comm-expan}, and since
$L_{e_K}$ is idempotent. Then, using the factorization property
$f = h \Mop k$ once more, we obtain
\begin{equation}
\bigl|\Trw\bigl((L_f - L_{f_K})\, (1+|\Dslash|)^{-n}\bigr)\bigr|
\leq \|L_h - L_{e_K \Mop h}\| \,
\bigl| \Trw\bigl( L_k\, (1 + |\Dslash|)^{-n} \bigr) \bigr|,
\label{eq:eval-trunc}
\end{equation}
and the right hand side vanishes as $K \to \infty$, thanks to the
estimate \eqref{eq:borne} for the norm of a twisted multiplication
operator. On rewriting \eqref{eq:res-value} as
$$
\Trw \bigl( L_{f_K}\, (1 + |\Dslash|)^{-n} \bigr)
= C(n) \int_M f_K(p) \,\mu_g(p),
$$
the left hand side converges to $\Trw(L_f\,(1 + |\Dslash|)^{-n})$
as $K \to \infty$. On the right hand side, the rapid decrease of the
coefficients $c_{mn}(\hat x)$ in \eqref{eq:f-trunc} ensures that
$f_K \to f$ in $L^1(M,\mu_g)$. Taking the limit as $K \to \infty$
on both sides of~\eqref{eq:eval-trunc} therefore yields the desired 
Dixmier trace evaluation:
$$
\Trw \bigl( L_f\, (1 + |\Dslash|)^{-n} \bigr)
= C(n) \int_M f(p) \,\mu_g(p).
\eqno \qed
$$
\hideqed
\end{proof}

\subsection*{Acknowledgments}

We thank Gilles Carron, Thierry Coulhon, Pierre Duclos,
Jos\'e M. Gracia-Bond\'{\i}a, Christophe Pittet, Adam Rennie, Thomas
Sch\"ucker, Andrei Teleman, Antony Wassermann and the referee for
helpful discussions and remarks. JCV is grateful for support from the
CNRS and the Vicerrector\'{\i}a de Investigaci\'on of the Universidad
de Costa Rica; and he thanks Piotr M. Hajac for his hospitality at the
University of Warsaw during the final stages of this work.

\end{document}